\documentclass[journal,10pt]{IEEEtran}

\usepackage{amsmath,amssymb}
\usepackage{subfigure}
\usepackage{graphicx,graphics,color,psfrag}
\usepackage{cite,balance}
\usepackage{algorithm}
\usepackage{accents}
\usepackage{amsthm}
\usepackage{bm}
\usepackage{url}
\usepackage{algorithmic}
\usepackage[english]{babel}
\usepackage{multirow}
\usepackage{enumerate}
\usepackage{cases}
\usepackage{stfloats}
\usepackage{dsfont}
\usepackage{color,soul}
\usepackage{amsfonts}
\usepackage{tcolorbox}

\usepackage{cite,graphicx,amsmath,amssymb}
\usepackage{fancyhdr}
\usepackage{hhline}
\usepackage{graphicx,graphics}
\usepackage{array,color}
\usepackage{amsmath}
\usepackage{amsthm}
\usepackage[flushleft]{threeparttable}
\IEEEoverridecommandlockouts

\newtheorem{proposition}{Proposition}

\newtheorem{lemma}{Lemma}

\include{header}

\newcommand{\mv}[1]{\mbox{\boldmath{$ #1 $}}}

\newcommand{\br}[1]{\bm{\mathrm{#1}}} 
\newcommand{\mr}[1]{\mathrm{#1}}

\makeatletter
\def\endthebibliography{%
	\def\@noitemerr{\@latex@warning{Empty `thebibliography' environment}}%
	\endlist
}
\makeatother

\begin{document}
	\title{Near-Field Integrated Sensing and Communication with Extremely Large-Scale Antenna Array} 
	\author{Haocheng Hua, \textit{Student Member, IEEE}, Jie Xu, \textit{Senior Member, IEEE}, 
		and Rui Zhang, \textit{Fellow, IEEE} \\
		\thanks{
			H. Hua and J. Xu are with the School of Science and Engineering and the Shenzhen Future Network of Intelligence Institute (FNii-Shenzhen), The Chinese University of Hong Kong (Shenzhen), Guandong 518172, China (e-mail: haochenghua@link.cuhk.edu.cn, xujie@cuhk.edu.cn). J. Xu is the corresponding author.}
		\thanks{R. Zhang is with the School of Science and Engineering, Shenzhen Research Institute of Big Data, The Chinese University of Hong Kong (Shenzhen), Guangdong 518172, China. He is also with the Department of Electrical and Computer Engineering, National University of Singapore, Singapore 117583 (e-mail: elezhang@nus.edu.sg).}
	}
	
	\maketitle
	
	\begin{abstract}
		This paper studies a near-field integrated sensing and communication (ISAC) system with extremely large-scale antenna array (ELAA), in which a base station (BS) deployed with a very large number of antennas transmits wireless signals to communicate with multiple communication users (CUs) and simultaneously uses the echo signals to localize multiple point targets in the three-dimension (3D) space. To balance the performance tradeoff between near-field communication and 3D target localization, we design the transmit covariance matrix at the BS to optimize the localization performance while ensuring the signal-to-interference-plus-noise ratio (SINR) constraints at individual CUs. In particular, we formulate three design problems by considering different 3D localization performance metrics, including minimizing the sum Cram\'er-Rao bound (CRB) for estimating 3D locations, maximizing the minimum target illumination power, and maximizing the minimum target echo signal power. Although the three design problems are non-convex in general, we obtain their global optimal solutions via the technique of semi-definite relaxation (SDR) by proving the tightness of such relaxations. It is rigorously shown that the optimal solutions to the three problems have low-rank structures depending on the sensing and communication channel matrices, which can be exploited to greatly reduce the computational complexity of the SDR-based solutions. Interestingly, we find that in the special case with a single collocated target/CU present towards the middle of a symmetric uniform planar array (UPA), the optimal solutions to the three problems become identical to the SINR-maximization design and have a closed form, while in other cases they can be different in general. Besides, when the target/CU moves away from the transmitter/receiver, the CRB may first decrease and then increase. These two phenomena differ from those in the far-field scenario. Numerical results show the benefits of the proposed designs in optimizing both sensing and communication performance, by exploiting the beam focusing capabilities of ELAA. In particular, with ELAA, sensing performance is significantly improved with negligible communication performance loss.
	\end{abstract}
	\begin{IEEEkeywords}
		Integrated sensing and communication (ISAC), 3D near-field localization, Cram\'er-Rao bound (CRB), extremely large-scale antenna array (ELAA), beamforming. 
	\end{IEEEkeywords}
	
	
	\section{Introduction} 
	\label{sec:intro}
	
	Integrated sensing and communication (ISAC) \cite{IMT_2030_6G_vision_new,liu2022integrated, hua2023optimal,hua2022mimojournal} has emerged as one of the most important scenarios and features of sixth-generation (6G) wireless networks, in which wireless communication infrastructures and spectrum resources are reused for radar sensing.
	The dual-functional ISAC systems can provide new services with significantly enhanced spectrum and cost efficiency, which have broad applications in low-altitude economy, smart factory, and vehicles-to-everything (V2X), etc. 
	Along with ISAC, 6G also exploits emerging technologies such as extremely large-scale antenna array (ELAA) and millimeter wave (mmWave)/terahertz (THz) \cite{heath2016overview} to enhance the performance of both wireless communications and radar sensing. As such, the large number of antennas and the high frequency band introduce a paradigm shift from the conventional far-field sensing and communication design considering the planar electromagnetic (EM) wavefront to the new near-field design, in which both functions are implemented in the near-field (or the Fresnel region \cite{zhang2022beam, hua2023near,cong2023near,wang2024tutorial}) and thus the spherical EM wavefront should be adopted \cite{lu2021communicating,lu2023near,zhang2022beam, cui2022channel, d2022cramer,hua2023near,guerra2021near}. 
	
	The near-field communications and sensing have been independently investigated in the literature. On one hand, various prior works investigated the near-field communications from different perspectives, such as the signal-to-noise ratio (SNR) scaling laws for single-user communications \cite{lu2021communicating}, beam focusing for multi-user communications \cite{zhang2022beam}, new channel estimation designs exploiting the polar-domain sparsity \cite{cui2022channel} and distance-parameterized angular-domain sparsity \cite{zhang2024near}, as well as near-field non-orthogonal multiple access (NOMA) designs \cite{ding2023resolution}. 
	On the other hand, there have been rich literature studying near-field sensing. For example, \cite{huang1991near} first proposed to exploit the spherical wavefront for target localization, and then \cite{lee1995covariance} proposed various algorithms for localizing near-field targets in the two-dimensional (2D) space. Besides, \cite{el2010conditional,grosicki2005weighted,gazzah2014crb} derived the fundamental performance limit of localization in terms of the Cram\'er-Rao bound (CRB), which characterizes the variance lower bound of any unbiased estimator \cite{levy2008principles}.
	Furthermore, some other works studied the localization of near-field targets in the 3D space, which is more practical in future wireless sensing scenarios like low-altitude economy. In this regard, 
	\cite{khamidullina2021conditional} derived the CRB for multi-target 3D localization by considering the case with orthogonal transmit waveforms and matched filtering reception. In addition, to facilitate transmit waveform adaptation, \cite{hua2023near} further considered a more general multi-target setup with generic transmit signal waveforms, exact spherical wavefronts, and general clutter models, based on which the corresponding localization CRBs were derived and efficient localization algorithms were proposed. 
	
	The research on near-field ISAC is still at its infant phase \cite{cong2023near,li2024near,wang2023near,zhao2024modeling,wang2024wideband}.
	Specifically, \cite{li2024near} and \cite{wang2023near} considered the target localization in the near-field ISAC systems and proposed beamforming designs to optimize CRB for target localization while meeting the communication requirements. However, these works mainly focused on localizing a single target in the 2D space by considering a uniform linear array (ULA) equipped at the base station (BS). While \cite{zhao2024modeling} developed various design strategies for 3D near-field ISAC systems under both the downlink and uplink scenarios, the considered setup only includes a single user and a single target. 
	More recently, \cite{wang2024wideband} studied the wideband orthogonal frequency division multiplexing (OFDM)-based near-field ISAC systems, in which the precoding matrix and the antenna selection strategies were optimized to balance the performance tradeoff between sensing and communication.

	Despite the above progress, there still remain various challenges in designing near-field ISAC systems with ELAAs, which require a further study. For example, ELAAs generally incorporate thousands of antennas at the BS, which pose great challenges on efficient beamforming design, while this issue has not been addressed yet in prior work. Besides, it still remains unclear how near-field ISAC designs with ELAA fundamentally differ from those in the far-field scenario, which deserves more in-depth investigation. These design challenges and open problems thus motivate this work.
	
	Specifically, we consider a BS equipped with ELAAs aiming to simultaneously localize multiple near-field targets in the 3D space and communicate with multiple communication users (CUs) via transmitting jointly designed information and sensing signals in the downlink. We adopt exact near-field wavefront modeling with both phase and channel amplitude variations. 	
	Our main results are given as follows. 
	\begin{itemize} 
		\item To balance the performance tradeoff, we design the transmit covariance matrix at the BS to optimize the localization performance while ensuring the individual signal-to-interference-plus-noise ratio (SINR) constraints at each CU. Specifically, we formulate three design problems by considering various localization performance metrics, including minimizing the sum-CRB for estimating 3D locations, maximizing the minimum target illumination power, and maximizing the minimum target echo signal power. These design problems are all non-convex and thus difficult to be optimally solved in general.
		\item We obtain the global optimal solution to the three design problems via semi-definite relaxation (SDR) by proving the tightness of such relaxations. More importantly, we prove that their optimal solutions have low-rank structures depending on the sensing and communication channel matrices, which can be exploited to greatly reduce the computational complexity of the SDR-based solutions. 
		\item To reveal more insights, we study the special case with one target and one CU and characterize the complete communication-sensing performance tradeoff curve by pinpointing two end points and revealing their corresponding design strategies. We further analyze the special scenario where the target is collocated with the CU. It is found that when the target/CU is present towards the middle of a symmetric uniform planar array (UPA), the optimal solutions to the proposed three designs become identical to the SINR-maximization design and have a simple closed form.
		\item Numerical results validate the effectiveness of proposed methods, particularly by utilizing the revealed low-rank structure to find the optimal solution.
		It is shown that in the collocated target/CU case, except for the aforementioned special configurations, the optimal strategies for CRB-minimization and SINR-maximization are different in general. Besides, when the target/CU moves away from the transmitter/receiver, the CRB may first decrease and then increase. These two findings are unique to the near-field setup and unavailable in the conventional far-field scenario. Moreover, thanks to ELAAs, all the proposed designs demonstrate the beam focusing capability and the distance-domain spatial multiplexing, which significantly enhance the performance of both near-field multi-user communication and multi-target 3D localization.
		Finally, it is revealed that with larger number of antennas and the exploitation of additional scattered non-line-of-sight (NLoS) paths from the target to the CU, better localization performance as well as communication performance can be achieved.
	\end{itemize}
	
	{\it Notations:} Boldface letters refer to vectors (lower case) or matrices (upper case). For a square matrix $\mv{M}$, ${\operatorname{tr}}(\mv{M})$, $\bm{M}^{-1}$, and $|\bm{M}|$ denote its trace, inverse, and determinant, respectively. For an arbitrary-size matrix $\mv{M}$, $\mathfrak{R}(\bm{M})$, $\mathfrak{I}(\bm{M})$, $\bm{M}^H$, $\bm{M}^*$, $\bm{M}^T$, $\bm{M}[m:n,p:q]$, and $\operatorname{vec}(\bm{M})$ denote its real part, imaginary part, conjugate transpose, conjugate, transpose, corresponding sub-block matrix with dimension $(n-m+1) \times (q-p+1)$, and vectorization, respectively, and $\odot$ is the Hadamard product.
	We use $\mathbb{R}^{x\times y}$ and $\mathbb{C}^{x\times y}$ to represent the spaces of real and complex matrices with dimension $x \times y$, respectively. {${\mathbb{E}}\{\cdot\}$} denotes the statistical expectation. $\|\mv{x}\|$ is the Euclidean norm of a complex vector $\mv{x}$ and $\operatorname{diag}(\bm{x})$ denotes a diagonal matrix with diagonal elements $\bm{x}$. The imaginary unit is written as $\mathrm{j} = \sqrt{-1}$ .
	
	\begin{figure}[t]
		\centering
		\includegraphics[width=3.3in]{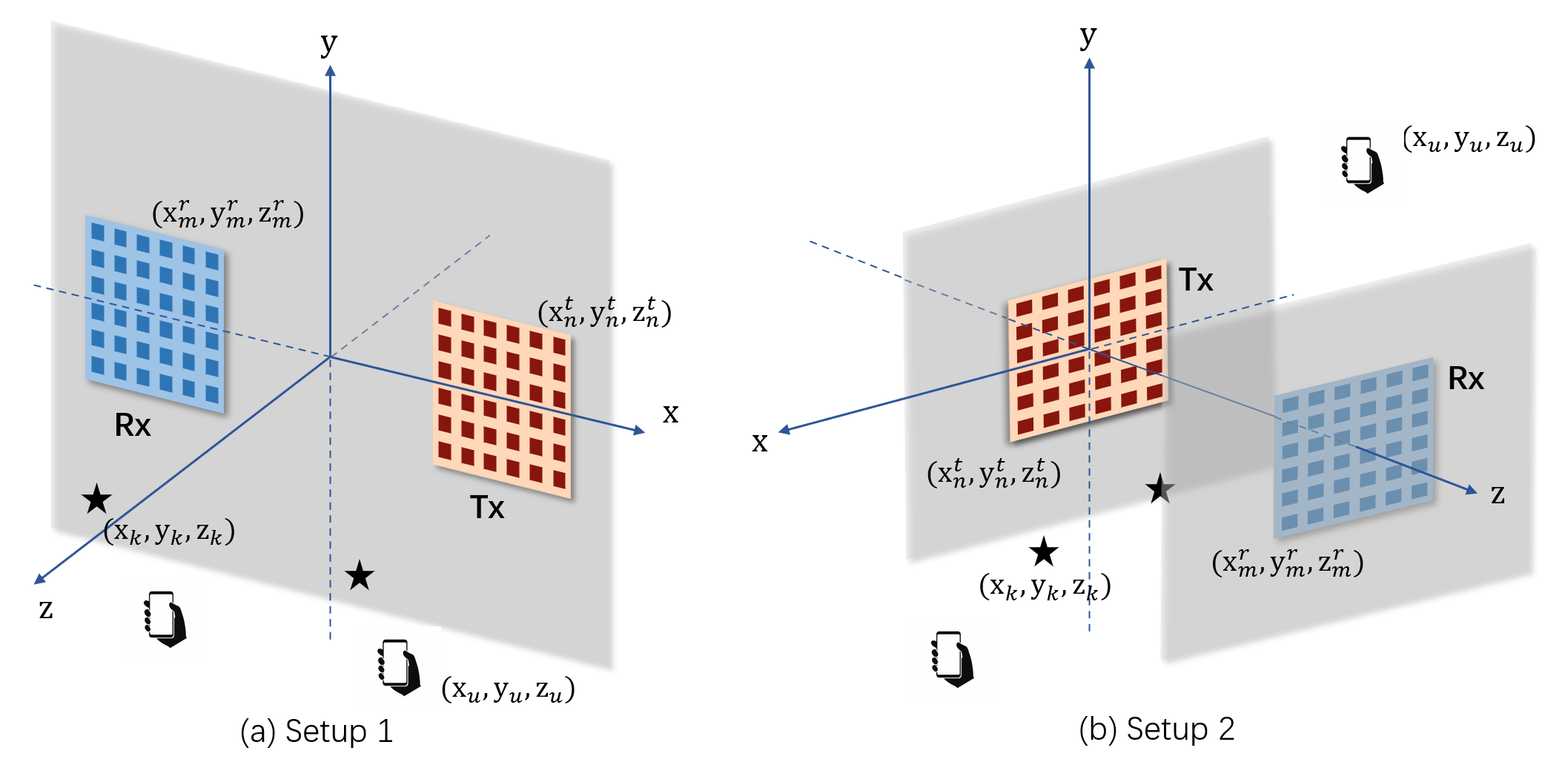}
		\centering
		\caption{Typical ISAC scenarios with multiple targets and CUs. Black stars denote the targets while phone icons denote the CUs.}
		\label{fig:sm_isac}
	\end{figure}
	
	\section{System Model}\label{Section_system_model}
	
	
	We consider two typical ISAC systems with ELAAs as shown in Fig. \ref{fig:sm_isac}, each of which consists of an ISAC transmitter (Tx) with $N$ antennas and a sensing receiver (Rx) with $M$ antennas, where $M$ and $N$ are in the order of thousands.
	Let $\br{l}^t_{n}  = (\mr{x}^t_n,\mathrm{y}^t_n,\mathrm{z}^t_n)$ and $\bm{\mathrm{l}}^r_{m} = (\mr{x}^r_m,\mathrm{y}^r_m,\mathrm{z}^r_m)$ denote the position of the $n$-th antenna at Tx and that of the $m$-th antenna at Rx, respectively, where $n \in \mathcal{N} \triangleq \{1,2,...,N\}$ and $m \in \mathcal{M} \triangleq \{1,2,...,M\}$.
	We consider that there are $K$ targets and $U$ CUs each equipped with a single antenna. Both targets and CUs are assumed to be located in the near-field region of the Tx/Rx ELAA. Let $\mathcal{K} \triangleq \{1,...,K\}$ denote the set of targets and $\bm{\mathrm{l}}_{k} = (\mathrm{x}_{k},\mathrm{y}_{k},\mathrm{z}_{k})$ the position of the $k$-th target. Let $\mathcal{U} \triangleq \{1,2,...,U\}$ denote the set of CUs and  $\br{l}_{u} = (\mr{x}_u,\mr{y}_u,\mr{z}_u)$ the position of the $u$-th CU.


	Let $L$ denote the total number of symbols of one sensing period. To facilitate ISAC, the BS transmits both information signals and dedicated sensing signals to perform multi-user downlink communication and 3D near-field target localization. Let $\bm{W}_C = \left[\bm{w}_1,...,\bm{w}_U\right] \in \mathbb{C}^{N \times U}$ and $\bm{W}_R = \left[\bm{w}^r_1,...,\bm{w}^r_N\right] \in \mathbb{C}^{N \times N}$ denote the digital beamformers for the CUs and dedicated radar digital beamformers, respectively. Let $\bm{s}_l = \left[s_{1,l},...,s_{U,l}\right]^T \in \mathbb{C}^{U}$ and $\bm{s}^r_l = \left[s^r_{1,l},...,s^r_{N,l}\right]^T \in \mathbb{C}^{N}$ denote the $l$-th transmitted communication symbol vector for all the $U$ users and the $l$-th transmitted dedicated radar symbol vector, respectively.
	The transmit signal vector of the BS is given by
	\begin{align}
		\bm{x}_l = \sum_{u=1}^{U} \bm{w}_u s_{u,l} + \sum_{n=1}^{N} \bm{w}_n^r s^r_{n,l}, \enspace \forall l \in \{1,...,L\},
	\end{align}
	where $\{s_{u,l}\}$ and $\{s^r_{n,l}\}$ are both independent random variables, mutually independent from each other, and with zero mean and unit variance. 
	%
	Let $\bm{R}_X \triangleq \mathbb{E}\{\bm{x}_l \bm{x}_l^H\}$ denote the covariance matrix of the transmit symbols. Assuming that $L$ is sufficiently large \cite{hua2022mimojournal,liu2021cramer}, it follows that
	\begin{align}
		\bm{R}_X = \frac{1}{L} \bm{X} \bm{X}^H = \bm{W}_C \bm{W}_C^H + \bm{R}_d,
	\end{align}
where $\bm{X} = \left[ \bm{x}_1, ..., \bm{x}_L \right]$ and $\bm{R}_d = \bm{W}_R \bm{W}_R^H$. 
As such, $\bm{W}_C$ and $\bm{R}_d$ are the design variables to be optimized. Once we find $\bm{R}_d$, we can perform the eigen-value decomposition (EVD) to retrieve all the dedicated sensing digital beamformers, i.e., $\left[\bm{w}_1^r,...,\bm{w}_N^r\right]$.

\subsection{Communication Performance Metric}

The received signal at the $u$-th CU at time $l$ is
\begin{align}
	y_{u,l} = \bm{h}_u^H \bm{x}_l + n_{u,l}, \enspace u \in \mathcal{U},
\end{align}
where $\bm{h}_u \in \mathbb{C}^{N}$ is the communication channel from the Tx to the $u$-th CU and $n_{u,l}$ is a circularly symmetric complex Gaussian (CSCG) noise term with zero mean and variance $\sigma_c^2$.
Accordingly, the SINR at each CU is
\begin{align}\label{equ:SINR_Type_I}
	\gamma_{u} (\{\bm{w}_u\}, \bm{R}_d) =\frac{\left|\bm{h}_u^{H} \bm{w}_u\right|^{2}}{\sum\limits_{k \in \mathcal{U} \atop k \neq u}\left|\bm{h}_u^{H} \bm{w}_k\right|^{2} + \bm{h}_u^H \bm{R}_d \bm{h}_u +\sigma_c^2}, \enspace u \in \mathcal{U}.
\end{align}
We assume that $\{\bm{h}_u\}$ is perfectly known at the BS via proper channel estimation mechanisms.

\subsection{Sensing Performance Metrics}

Assuming the narrowband sensing, the received echo signals over the $L$ symbols are denoted as $\bm{Y}_R \in \mathbb{C}^{M \times L}$, which is given by
\begin{align}\label{equ:Rx_data_matrix_isac}
	\bm{Y}_R = \sum_{k = 1}^{K} b_k \bm{a}(\bm{\mathrm{l}}_k)  \bm{v}^T(\bm{\mathrm{l}}_k) \bm{X} + \bm{Z}_R,
\end{align}
where $\bm{Z}_R = \left[\bm{z}_1,...,\bm{z}_L\right] \in \mathbb{C}^{M \times L}$ denotes the noise and interference with each column being independent and identically distributed (i.i.d.) CSCG random vectors with zero mean and covariance $\bm{Q} \in \mathbb{C}^{M \times M}$, and $b_k$ denotes the target complex reflection coefficient
of the $k$-th target. 
Furthermore, in (\ref{equ:Rx_data_matrix_isac}),
$\bm{a}(\bm{\mathrm{l}}_k)$ and $\bm{v}(\bm{\mathrm{l}}_k)$ denote the steering vectors at the Rx and Tx for the $k$-th target, respectively, given by 
\begin{align}\label{equ:steer_Rx_isac}
	\bm{a}(\bm{\mathrm{l}}_k) = [\alpha_1^r(\bm{\mathrm{l}}_k)  e^{-\mathrm{j} \nu \|\bm{\mathrm{l}}^r_1 - \bm{\mathrm{l}}_{k}\|},..., \alpha_{M}^r(\bm{\mathrm{l}}_k) e^{-\mathrm{j} \nu \|\bm{\mathrm{l}}^r_M - \bm{\mathrm{l}}_{k}\|} ]^T,
\end{align}
\begin{align}\label{equ:steer_Tx_isac}
	\bm{v}(\bm{\mathrm{l}}_k) = [\alpha_1^t(\bm{\mathrm{l}}_k)e^{-\mathrm{j} \nu \|\bm{\mathrm{l}}^t_1 - \bm{\mathrm{l}}_{k}\|},..., \alpha_N^t(\bm{\mathrm{l}}_k)e^{-\mathrm{j} \nu \|\bm{\mathrm{l}}^t_N - \bm{\mathrm{l}}_{k}\|} ]^T,
\end{align}
where $\nu = \frac{2 \pi}{\lambda}$ is the wave number of the carrier, $\lambda$ is the wavelength, $\alpha_m^r(\bm{\mathrm{l}}_k) = \frac{\lambda}{4 \pi \|\bm{\mathrm{l}}_k - \bm{\mathrm{l}}^r_m\|}$ denotes the distance-dependent channel amplitude from the $k$-th target to the $m$-th receive antenna based on the free-space path-loss model, and $\alpha_n^t(\bm{\mathrm{l}}_k) = \frac{\lambda}{4 \pi \|\bm{\mathrm{l}}_k - \bm{\mathrm{l}}^t_{n}\|}$ denotes that from the $k$-th target to the $n$-th transmit antenna. Note that in (\ref{equ:Rx_data_matrix_isac}), we assume that the CUs are small such that their effects on the received echo waves are negligible, and in (\ref{equ:steer_Rx_isac}) and (\ref{equ:steer_Tx_isac}), we assume omni-directional antennas with unit antenna gains. For ease of exposition, define
\begin{align}
	\label{equ:A_compact_isac}
	& \bm{A}(\{\bm{\mathrm{l}}_k\}) = \left[\bm{a}(\bm{\mathrm{l}}_1),\bm{a}(\bm{\mathrm{l}}_2),...,\bm{a}(\bm{\mathrm{l}}_K)\right] \in \mathbb{C}^{M \times K}, \\
	\label{equ:V_compact_isac}
	& \bm{V}(\{\bm{\mathrm{l}}_k\})  = \left[\bm{v}(\bm{\mathrm{l}}_1),\bm{v}(\bm{\mathrm{l}}_2),...,\bm{v}(\bm{\mathrm{l}}_K)\right] \in \mathbb{C}^{N \times K}, \\
	\label{equ:b_complete_RI_isac}
	& \bm{b}  = \left[b_1,...,b_K\right]^T =  \left[b_{\text{R}_1}+ \mathrm{j}b_{\text{I}_1},...,b_{\text{R}_K}+\mathrm{j}b_{\text{I}_K}\right]^T, \\
	& \bm{B}  = \operatorname{diag}(\bm{b}),
\end{align}
where $b_{\text{R}_k}$ and $b_{\text{I}_k}$ denote the real and imaginary parts of $b_k$, respectively. In (\ref{equ:Rx_data_matrix_isac}), the unknown parameters include the 3D target locations $\{\bm{\mathrm{l}}_k\}$, complex reflection coefficients $\bm{b}$, and the noise and interference covariance matrix $\bm{Q}$, while the BS aims to estimate $\{\br{l}_k\}$.
In the following, we consider three different metrics for sensing, namely CRB, target illumination power, and target echo signal power, respectively. 
While CRB is a more explicit theoretical metric characterizing the performance limit of localization, the illumination or echo signal power is a more general one that can be practically measured easily and is widely adopted in both localization and detection tasks.
\subsubsection{Metric I - CRB}
Let $\bm{\theta}$ denote a vector containing all the unknown target-related parameters in (\ref{equ:Rx_data_matrix_isac}), defined as
\begin{align}\label{eq:theta_interested_isac}
	\nonumber
	\bm{\theta} & = \left[\mathrm{x}_1,...,\mathrm{x}_K,\mathrm{y}_1,...,\mathrm{y}_K,\mathrm{z}_1,...,\mathrm{z}_K,b_{\text{R}_1},..,b_{\text{R}_K},b_{\text{I}_1},...,b_{\text{I}_K}\right]^T \\
	& = \left[\bm{\mathrm{x}},\bm{\mathrm{y}},\bm{\mathrm{z}},\bm{b}_\text{R},\bm{b}_\text{I}\right]^T \in \mathbb{R}^{5K}.
\end{align}
The Fisher information matrix (FIM) $\br{F}$ with respect to $\bm{\theta}$ is derived in \cite{hua2023near} and is shown below for exposition purpose. For simplicity, we omit $\{\br{l}_k\}$ in (\ref{equ:A_compact_isac}) and (\ref{equ:V_compact_isac}) in the following.
\begin{proposition}\label{Pro:F_deri}
	\emph{The FIM $\bm{\mathrm{F}} \in \mathbb{R}^{5K \times 5K}$ with respect to $\bm{\theta}$ is
		\begin{align}\label{eq:fisher_mat_sim}
			2 \left[
			\arraycolsep=0.75pt\def\arraystretch{1.2}
			\begin{array}{ccccc}
				\mathfrak{R}(\bm{\mathrm{F}}_{\bm{\mathrm{xx}}}) & \mathfrak{R}(\bm{\mathrm{F}}_{\bm{\mathrm{xy}}}) & \mathfrak{R}(\bm{\mathrm{F}}_{\bm{\mathrm{xz}}}) & \mathfrak{R}(\bm{\mathrm{F}}_{\bm{\mathrm{xb}}}) & -\mathfrak{I}(\bm{\mathrm{F}}_{\bm{\mathrm{xb}}}) \\[1pt]
				\mathfrak{R}(\bm{\mathrm{F}}^T_{\bm{\mathrm{xy}}}) & \mathfrak{R}(\bm{\mathrm{F}}_{\bm{\mathrm{yy}}}) & \mathfrak{R}(\bm{\mathrm{F}}_{\bm{\mathrm{yz}}}) & \mathfrak{R}(\bm{\mathrm{F}}_{\bm{\mathrm{yb}}}) & -\mathfrak{I}(\bm{\mathrm{F}}_{\bm{\mathrm{yb}}}) \\[1pt]
				\mathfrak{R}(\bm{\mathrm{F}}^T_{\bm{\mathrm{xz}}}) & \mathfrak{R}(\bm{\mathrm{F}}^T_{\bm{\mathrm{yz}}}) & \mathfrak{R}(\bm{\mathrm{F}}_{\bm{\mathrm{zz}}}) & \mathfrak{R}(\bm{\mathrm{F}}_{\bm{\mathrm{zb}}}) & -\mathfrak{I}(\bm{\mathrm{F}}_{\bm{\mathrm{zb}}}) \\[1pt]
				\mathfrak{R}(\bm{\mathrm{F}}^T_{\bm{\mathrm{xb}}}) & \mathfrak{R}(\bm{\mathrm{F}}^T_{\bm{\mathrm{yb}}}) & \mathfrak{R}(\bm{\mathrm{F}}^T_{\bm{\mathrm{zb}}}) & \mathfrak{R}(\bm{\mathrm{F}}_{\bm{\mathrm{bb}}}) & -\mathfrak{I}(\bm{\mathrm{F}}_{\bm{\mathrm{bb}}}) \\[1pt]
				-\mathfrak{I}(\bm{\mathrm{F}}^T_{\bm{\mathrm{xb}}}) & -\mathfrak{I}(\bm{\mathrm{F}}^T_{\bm{\mathrm{yb}}}) & -\mathfrak{I}(\bm{\mathrm{F}}^T_{\bm{\mathrm{zb}}}) & -\mathfrak{I}(\bm{\mathrm{F}}^T_{\bm{\mathrm{bb}}}) & \mathfrak{R}(\bm{\mathrm{F}}_{\bm{\mathrm{bb}}})
			\end{array}\right], 
		\end{align}
		where $\bm{\mathrm{F}}_{\bm{\mathrm{xx}}}, \bm{\mathrm{F}}_{\bm{\mathrm{yy}}}$, and $\bm{\mathrm{F}}_{\bm{\mathrm{zz}}}$ are given by
		\begin{align}\label{eq:F_multiple_list_xx_yy_zz}
			\bm{\mathrm{F}}_{\bm{\mathrm{uu}}} & =  L (\dot{\bm{A}}_{\bm{\mathrm{u}}}^H  \bm{Q}^{-1} \dot{\bm{A}_{\bm{\mathrm{u}}}}) \odot (\bm{B}^* \bm{V}^H \bm{R}_X^* \bm{V} \bm{B}) \\
			\nonumber
			& + L (\dot{\bm{A}}_{\bm{\mathrm{u}}}^H  \bm{Q}^{-1} \bm{A}) \odot (\bm{B}^* \bm{V}^H \bm{R}_X^* \dot{\bm{V}}_{\bm{\mathrm{u}}} \bm{B}) \\
			\nonumber
			& + L (\bm{A}^H  \bm{Q}^{-1} \dot{\bm{A}_{\bm{\mathrm{u}}}}) \odot (\bm{B}^* \dot{\bm{V}}_{\bm{\mathrm{u}}}^H \bm{R}_X^* \bm{V} \bm{B})\\
			\nonumber
			& + L (\bm{A}^H  \bm{Q}^{-1} \bm{A}) \odot (\bm{B}^* \dot{\bm{V}}_{\bm{\mathrm{u}}}^H \bm{R}_X^* \dot{\bm{V}}_{\bm{\mathrm{u}}} \bm{B}), \enspace \bm{\mathrm{u}} \in \{\bm{\mathrm{x}},\bm{\mathrm{y}},\bm{\mathrm{z}}\}, 
		\end{align}
		$\bm{\mathrm{F}}_{\bm{\mathrm{xy}}}, \bm{\mathrm{F}}_{\bm{\mathrm{xz}}}$, and $\bm{\mathrm{F}}_{\bm{\mathrm{yz}}}$ are given by
		\begin{align}\label{eq:F_multiple_list_xy_xz_yz}
			& \bm{\mathrm{F}}_{\bm{\mathrm{uv}}}  =  L (\dot{\bm{A}}_{\bm{\mathrm{u}}}^H  \bm{Q}^{-1} \dot{\bm{A}_{\bm{\mathrm{v}}}}) \odot (\bm{B}^* \bm{V}^H \bm{R}_X^* \bm{V} \bm{B}) \\
			\nonumber
			& + L (\dot{\bm{A}}_{\bm{\mathrm{u}}}^H  \bm{Q}^{-1} \bm{A}) \odot (\bm{B}^* \bm{V}^H \bm{R}_X^* \dot{\bm{V}}_{\bm{\mathrm{v}}} \bm{B}) \\
			\nonumber
			& + L (\bm{A}^H  \bm{Q}^{-1} \dot{\bm{A}_{\bm{\mathrm{v}}}}) \odot (\bm{B}^* \dot{\bm{V}}_{\bm{\mathrm{u}}}^H \bm{R}_X^* \bm{V} \bm{B})\\
			\nonumber
			& + L (\bm{A}^H  \bm{Q}^{-1} \bm{A}) \odot (\bm{B}^* \dot{\bm{V}}_{\bm{\mathrm{u}}}^H \bm{R}_X^* \dot{\bm{V}}_{\bm{\mathrm{v}}} \bm{B}), \enspace \bm{\mathrm{uv}} \in \{\bm{\mathrm{xy}},\bm{\mathrm{xz}},\bm{\mathrm{yz}}\}, 
		\end{align}
		and $\bm{\mathrm{F}}_{\bm{\mathrm{bb}}}$, $\bm{\mathrm{F}}_{\bm{\mathrm{xb}}}, \bm{\mathrm{F}}_{\bm{\mathrm{yb}}}$, and $\bm{\mathrm{F}}_{\bm{\mathrm{zb}}}$ are given by
		\begin{align}\label{eq:F_multiple_list_xb_yb_zb}
			\bm{\mathrm{F}}_{\bm{\mathrm{bb}}} & = L (\bm{A}^H  \bm{Q}^{-1} \bm{A}) \odot (\bm{V}^H \bm{R}_X^* \bm{V}),\\
			\bm{\mathrm{F}}_{\bm{\mathrm{ub}}}  & =  L (\dot{\bm{A}}_{\bm{\mathrm{u}}}^H  \bm{Q}^{-1} \bm{A}) \odot (\bm{B}^* \bm{V}^H \bm{R}_X^* \bm{V}) \\
			\nonumber
			& + L (\bm{A}^H  \bm{Q}^{-1} \bm{A}) \odot (\bm{B}^* \dot{\bm{V}}_{\bm{\mathrm{u}}}^H \bm{R}_X^* \bm{V}), \enspace \bm{\mathrm{u}} \in \{\bm{\mathrm{x}},\bm{\mathrm{y}},\bm{\mathrm{z}}\}, 
		\end{align}
		respectively. Here,
		\begin{align}\label{eq:VX_partial}
			\dot{\bm{A}_{\bm{\mathrm{u}}}}  = & \left[\frac{\partial \bm{a}(\bm{\mathrm{l}}_1)}{\partial \mathrm{u}_1},...,\frac{\partial \bm{a}(\bm{\mathrm{l}}_K)}{\partial \mathrm{u}_K}\right], \enspace \bm{\mathrm{u}} \in \{\bm{\mathrm{x}},\bm{\mathrm{y}},\bm{\mathrm{z}}\},   \\
			\dot{\bm{V}_{\bm{\mathrm{u}}}}  = & \left[\frac{\partial \bm{v}(\bm{\mathrm{l}}_1)}{\partial \mathrm{u}_1},...,\frac{\partial \bm{v}(\bm{\mathrm{l}}_K)}{\partial \mathrm{u}_K}\right], \enspace \bm{\mathrm{u}} \in \{\bm{\mathrm{x}},\bm{\mathrm{y}},\bm{\mathrm{z}}\}, \\
			\label{eq:AX_partial_element}
			\left(\frac{\partial \bm{a}(\bm{\mathrm{l}}_k)}{\partial \mathrm{u}_k}\right)_m & = \bm{a}_m(\bm{\mathrm{l}}_k)(\frac{\mathrm{u}_m^r-\mathrm{u}_k}{\|\bm{\mathrm{l}}^r_m - \bm{\mathrm{l}}_{k}\|^2}+\mathrm{j} \nu \frac{\mathrm{u}_m^r-\mathrm{u}_k}{\|\bm{\mathrm{l}}^r_m - \bm{\mathrm{l}}_{k}\|}),   \\
			\label{eq:VX_partial_element}
			\left(\frac{\partial \bm{v}(\bm{\mathrm{l}}_k)}{\partial \mathrm{u}_k}\right)_n & = \bm{v}_n(\bm{\mathrm{l}}_k)(\frac{\mathrm{u}_n^t-\mathrm{u}_k}{\|\bm{\mathrm{l}}^t_n - \bm{\mathrm{l}}_{k}\|^2}+\mathrm{j} \nu \frac{\mathrm{u}_n^t-\mathrm{u}_k}{\|\bm{\mathrm{l}}^t_n - \bm{\mathrm{l}}_{k}\|}). 
		\end{align}
		In (\ref{eq:AX_partial_element}) and (\ref{eq:VX_partial_element}), $\bm{a}_m(\bm{\mathrm{l}}_k)$ and $\bm{v}_n(\bm{\mathrm{l}}_k)$ denote the $m$-th element of $\bm{a}(\bm{\mathrm{l}}_k)$ and the $n$-th element of $\bm{v}(\bm{\mathrm{l}}_k)$, respectively.}
\end{proposition}
It is worth noting that $\br{F}$ is a function of $\bm{R}_X$. Accordingly, the CRB matrix $\bm{\mathrm{C}}$ for estimating $\bm{\theta}$ is
	$\bm{\mathrm{C}} = \bm{\mathrm{F}}^{-1},$
	and the CRB for localizing the $k$-th target is
	\begin{align}\label{eq:CRB_position_isac}
		\text{CRB}_k & =  \text{CRB}_{k,\mathrm{x}} + \text{CRB}_{k,\mathrm{y}} + \text{CRB}_{k,\mathrm{z}}\\
		\nonumber
		& = \bm{\mathrm{C}}[k,k] + \bm{\mathrm{C}}[k+K,k+K] + \bm{\mathrm{C}}[k+2K,k+2K],
	\end{align}
	where $\text{CRB}_{k,\mathrm{u}}, \mathrm{u} \in \{\mathrm{x},\mathrm{y},\mathrm{z}\}$, denotes the CRB for estimating the $\mathrm{u}$-coordinate of the $k$-th target. To balance the localization performance of all the targets, we aim to minimize the summation of the CRBs for localizing all the targets, i.e.,
	\begin{align}\label{eq:Sum_CRB}
		\text{CRB}_{\Sigma}(\bm{R}_X) = \sum_{k=1}^{K}\text{CRB}_k,
	\end{align}
	which is also a function of $\bm{R}_X$.

	\subsubsection{Metric II - Target Illumination Power} 
	Besides CRB, inspired by the conventional metric of transmit beampattern \cite{hua2023optimal,Eldar2020joint}, we further propose two alternative sensing performance metrics.
	The first one is target illumination power, which also corresponds to the 3D transmit beampattern gain, given as
	\begin{align}\label{eq:Tx_gain}
		E^t(\br{l}_k) & \triangleq  \mathbb{P}(\mathrm{x}_k,\mathrm{y}_k,\mathrm{z}_k) =  \mathbb{E}\left[\|\bm{v}^T(\br{l}_k)\bm{x}_l\|^2\right]\\
		\nonumber
		& = \bm{v}^T(\br{l}_k) \bm{R}_X \bm{v}^*(\br{l}_k), \enspace k \in \mathcal{K}.
	\end{align}
	Based on \textit{Metric II}, we aim to maximize the minimum target illumination power in (\ref{eq:Tx_gain}) over $k \in \mathcal{K}$.
	\subsubsection{Metric III - Target Echo Signal Power}
	The second one is the echo signal power from each target at the Rx, given by 
	\begin{align}\label{eq:Rx_gain}
		E^r(\br{l}_k) & = \mathbb{E}\left[\|\bm{a}(\br{l}_k)b_k\bm{v}^T(\br{l}_k)\bm{x}_l\|^2\right] \\
		\nonumber
		& = \|\bm{a}(\br{l}_k)\|^2 |b_k|  \bm{v}^T(\br{l}_k) \bm{R}_X \bm{v}^*(\br{l}_k), \enspace k \in \mathcal{K}.
	\end{align}
	Based on \textit{Metric III}, we aim to maximize the minimum target echo signal power in (\ref{eq:Rx_gain}) over $k \in \mathcal{K}$.
	
	Notice that in the above three sensing metrics, it is assumed that $\{\br{l}_k\}$, $\{b_k\}$, and $\bm{Q}$ are all obtained in the previous estimation phase and thus known at the BS in advance to facilitate the transmit design \cite{liu2021cramer, hua2022mimojournal}.

	\section{SINR-constrained Sum-CRB Minimization}\label{sec:CRB_min}
	
	
	
	This section considers the transmit design by employing \textit{Metric I} for sensing.
	The corresponding SINR-constrained sum-CRB minimization problem is formulated as
	\begin{subequations}
		\begin{align}\label{eq:CRB_ISAC_PI}
			(\text{P1}): & \min_{\{\bm{w}_u\}_{u=1}^U,\bm{R}_d\succeq \bm{0}, \bm{R}_X} \text{ } \sum_{k=1}^{K}\text{CRB}_k(\bm{R}_X) \\
			\label{eq:SINR}
			\text{s.t. } & \gamma_{u} (\{\bm{w}_u\}, \bm{R}_d) \geq \Gamma_u, \forall u \in \mathcal{U},\\
			\label{eq:Sum_power}
			&  \sum_{u=1}^U \|\bm{w}_u\|^2 +\operatorname{tr}(\bm{R}_d) \leq P_T,\\
			\label{eq:Rx}
			& \bm{R}_X = \bm{R}_d + \sum_{u=1}^{U} \bm{w}_u \bm{w}_u^H, 
		\end{align}
	\end{subequations}
	where $P_T$ is the total transmit power budget. 
	
	
	
	
	Although (P1) is non-convex, we can transform it into a semi-definite program (SDP) via SDR. 
	By introducing $3K$ auxiliary variables $\{t_k\}_{k=1}^{3K}$ satisfying $t_k - \br{e}_k^T \br{F}^{-1} \br{e}_k \geq 0, \forall k \in \{1,...,3K\}$,
	where $\br{e}_k$ is the $k$-th column of an identity matrix with dimension $5K \times 5K$,
	(P1) is transformed into
	\begin{subequations}
		\begin{align}\label{eq:CRB_sensing_SDP_pre}
			(\text{P1.1}): & \min_{\{t_k\}_{k=1}^{3K},\{\bm{w}_u\}_{u=1}^U,\bm{R}_d \succeq \bm{0}, \bm{R}_X} \text{ } \sum_{k=1}^{3K} t_k \\
			\label{eq:P1_1_CRB_scalar}
			\text{s.t. } & t_k - \br{e}_k^T \br{F}^{-1} \br{e}_k \geq 0, \enspace \forall k \in \{1,...,3K\}, \\
			&  \gamma_{u} (\{\bm{w}_u\}, \bm{R}_d) \geq \Gamma_u, \enspace \forall u \in \mathcal{U},\\
			&  \sum_{u=1}^U \|\bm{w}_u\|^2 +\operatorname{tr}(\bm{R}_d) \leq P_T, \\
			& \bm{R}_X = \bm{R}_d + \sum_{u=1}^{U} \bm{w}_u \bm{w}_u^H.
		\end{align}
	\end{subequations}
	In (P1.1), (\ref{eq:P1_1_CRB_scalar}) can be further transformed into a linear matrix inequality (LMI) constraint via the Schur complement. Besides, by defining $\bm{W}_u = \bm{w}_u \bm{w}_u^H, u \in \mathcal{U}$, (P1.1) is further transformed into 
	\begin{subequations}
		\begin{align}\label{eq:CRB_ISAC_P2_alter}
			(\text{P1.2}): & \min_{\{t_k\}_{k=1}^{3K}, \{\bm{W}_u\}_{u=1}^U,\bm{R}_d \succeq \bm{0}, \bm{R}_X} \text{ } \sum_{k=1}^{3K} t_k \\
			\label{eq:CRB_ISAC_P2r_Fish}
			\text{s.t. } & \left[
			\arraycolsep=0.75pt\def\arraystretch{1.2}
			\begin{array}{cc}
				\br{F} & \br{e}_k \\[1pt]
				\br{e}_k^T & t_k
			\end{array}\right] \succeq \bm{0}, \enspace \forall k \in \{1,...,3K\}, \\
			\label{eq:CRB_ISAC_P2r_SINR}
			& \frac{\text{tr}\left(\bm{h}_{u} \bm{h}_{u}^H \bm{W}_{u}\right)}{\Gamma_u}-\sum\limits_{k \in \mathcal{U} \atop k \neq u} \text{tr}\left(\bm{h}_u \bm{h}_u^H \bm{W}_{k}\right)\\
			\nonumber
			& \quad - \text{tr}\left(\bm{h}_u \bm{h}_u^H \bm{R}_{d}\right)-\sigma_{c}^{2} \geq 0, \enspace \forall u \in \mathcal{U},\\
			\label{eq:CRB_ISAC_P2r_power}
			&  \sum_{u=1}^U \operatorname{tr}(\bm{W}_u) +\operatorname{tr}(\bm{R}_d) \leq P_T, \\
			& \bm{R}_X = \bm{R}_d + \sum_{u=1}^{U} \bm{W}_u, \\
			\label{eq:rank_1_P21}
			& \bm{W}_u \succeq \bm{0}, \enspace \operatorname{rank}(\bm{W}_u) = 1,  \enspace \forall u \in \mathcal{U}.
		\end{align}
	\end{subequations}
	Removing the rank-one constraints in (\ref{eq:rank_1_P21}), (P1.2) is relaxed as (SDR1.2), which is a standard SDP.
	However, solving (SDR1.2) via the interior point method could incur a huge computational complexity, particularly with ELAAs equipped at Tx/Rx. Specifically, in (SDR1.2), the design variables contain $(U+1)$ matrices each with dimension $N \times N$. The formidable computational complexity in solving (SDR1.2) motivates us to exploit the solution structure of (SDR1.2). 
	
	
	Let $\bm{H}_c = \left[\bm{h}_1,...,\bm{h}_U\right]$, we have the following proposition.
	\begin{proposition}\label{Prop:Rx_span}
		\emph{For any global optimal solution to (SDR1.2), denoted as $\{\{\bm{W}_u\},\bm{R}_d, \bm{R}_X\}$, their conjugates $\bm{W}_u^*, u \in \mathcal{U}$, and $\bm{R}_d^*$ each have a maximum rank of $4K+U$ (assuming $4K+U \leq N$) with eigenvectors all belonging to the subspace spanned by the columns of $\bm{V}$, $\dot{\bm{V}}_{\br{x}}$, $\dot{\bm{V}}_{\br{y}}$, $\dot{\bm{V}}_{\br{z}}$, and $\bm{H}_c^*$, i.e., 
			\begin{align}
				\label{eq:Wp_low_rank}
				\bm{W}_u^* & = \br{U}_{sc} \br{\Sigma}_u \br{U}_{sc}^H, \enspace  u \in \mathcal{U}, \\
				\label{eq:Rd_low_rank}
				\bm{R}_d^* & = \br{U}_{sc} \br{\Sigma}_d \br{U}_{sc}^H.
			\end{align}
			Accordingly, the optimal $\bm{R}_X$ is written as
			\begin{align}
				\bm{R}_X = \br{U}_{sc}^* \br{\Sigma}_{sc}^* \br{U}_{sc}^T,
			\end{align}
			where $\br{\Sigma}_{sc} = \sum_{u=1}^{U}\br{\Sigma}_u + \br{\Sigma}_d$ is a $(4K+U) \times (4K+U)$ positive semi-definite matrix, and $\br{U}_{sc}$ is given in (\ref{eq:U_sc_expression}) at the top of this page.
			\begin{figure*}[t]
				\begin{align}
					\label{eq:U_sc_expression}
					\br{U}_{sc} = \left[ \arraycolsep=0.35pt\def\arraystretch{1.2}
					\bm{H}_c^* (\bm{H}_c^T \bm{H}_c^*)^{\text{-}\frac{1}{2}}  \enspace
					\bm{V} (\bm{V}^H \bm{V})^{\text{-}\frac{1}{2}}  \enspace \dot{\bm{V}}_{\br{x}} (\dot{\bm{V}}_{\br{x}}^H \dot{\bm{V}}_{\br{x}})^{\text{-}\frac{1}{2}} \enspace \dot{\bm{V}}_{\br{y}} (\dot{\bm{V}}_{\br{y}}^H \dot{\bm{V}}_{\br{y}})^{\text{-}\frac{1}{2}} \enspace \dot{\bm{V}}_{\br{z}} (\dot{\bm{V}}_{\br{z}}^H \dot{\bm{V}}_{\br{z}})^{\text{-}\frac{1}{2}} \right],
				\end{align}
				\begin{center}
					\rule{18.0 cm}{0.02 cm}
				\end{center}
			\end{figure*}
			\begin{proof}
				See Appendix \ref{app:low_rank_structure}.
		\end{proof}}
	\end{proposition}
	Let the QR decomposition of $\br{U}_{sc}$ be $\br{U}_{sc} = \br{Q}_{sc} \br{R}_{sc}$,
	with $\br{Q}_{sc}^H \br{Q}_{sc} = \bm{I}_{4K+U}$. Thus, (\ref{eq:Wp_low_rank}) and (\ref{eq:Rd_low_rank}) are rewritten as
	\begin{align}
		\label{eq:Wp_low_rank_qr}
		\bm{W}_u^* & = \br{Q}_{sc} \br{R}_{sc} \br{\Sigma}_u \br{R}_{sc}^H \br{Q}_{sc}^H, \enspace u \in \mathcal{U}, \\
		\label{eq:Rd_low_rank_qr}
		\bm{R}_d^* & = \br{Q}_{sc} \br{R}_{sc} \br{\Sigma}_d \br{R}_{sc}^H \br{Q}_{sc}^H.
	\end{align}
	Based on the facts that $\tilde{\br{\Sigma}}_u \triangleq \br{R}_{sc} \br{\Sigma}_u \br{R}_{sc}^H \succeq \bm{0}, \forall u \in \mathcal{U}$, $\tilde{\br{\Sigma}}_d \triangleq \br{R}_{sc} \br{\Sigma}_d \br{R}_{sc}^H \succeq \bm{0}$, $\tilde{\br{\Sigma}}_{sc} \triangleq \br{R}_{sc} \br{\Sigma}_{sc} \br{R}_{sc}^H \succeq \bm{0}$, and Proposition \ref{Prop:Rx_span}, (SDR1.2) is further transformed into 
	\begin{subequations}
		\begin{align}\label{eq:CRB_ISAC_P2R}
			(\text{SDR1.3}&):\min_{\{t_k\}_{k=1}^{3K},\{\tilde{\br{\Sigma}}_u\}_{u=1}^U,\tilde{\br{\Sigma}}_d\succeq\bm{0}, \tilde{\br{\Sigma}}_{sc}}  \sum_{k=1}^{3K} t_k \\
			\label{eq:CRB_ISAC_P2R_Fish}
			\text{s.t. } & \left[
			\arraycolsep=0.75pt\def\arraystretch{1.2}
			\begin{array}{cc}
				\br{F} & \br{e}_k \\[1pt]
				\br{e}_k^T & t_k
			\end{array}\right] \succeq \bm{0}, \enspace \forall k \in \{1,...,3K\}, \\
			\nonumber
			& \frac{\text{tr}\left(\bm{h}_u \bm{h}_u^H \br{Q}_{sc}^* \tilde{\br{\Sigma}}_u^* \br{Q}_{sc}^T \right)}{\Gamma_u}-\sum\limits_{k \in \mathcal{U} \atop k \neq u} \text{tr}\left(\bm{h}_u \bm{h}_u^H \br{Q}_{sc}^* \tilde{\br{\Sigma}}_k^* \br{Q}_{sc}^T\right) \\
			\label{eq:CRB_ISAC_P2R_SINR}
			&  - \text{tr}\left(\bm{h}_u \bm{h}_u^H \br{Q}_{sc}^* \tilde{\br{\Sigma}}_d^* \br{Q}_{sc}^T\right)-\sigma_{c}^{2} \geq 0, \enspace \forall u \in \mathcal{U},\\
			\label{eq:CRB_ISAC_P2R_Power}
			&  \sum_{u=1}^U \operatorname{tr}(\tilde{\br{\Sigma}}_u^*) +\operatorname{tr}(\tilde{\br{\Sigma}}_d^*) \leq P_T,\\
			& \tilde{\br{\Sigma}}_{sc} = \tilde{\br{\Sigma}}_d + \sum_{u=1}^{U}\tilde{\br{\Sigma}}_u, \enspace \tilde{\br{\Sigma}}_u \succeq \bm{0}, \enspace \forall u \in \mathcal{U}.
		\end{align}
	\end{subequations}
	Now, the design variables $\{\tilde{\br{\Sigma}}_u\}$ and $\tilde{\br{\Sigma}}_d$ only have dimension of $(4K+U) \times (4K+U)$, and as a result, the computational complexity is drastically reduced. We can then solve the SDP (SDR1.3) with reduced complexity and accordingly obtain the optimal solution $\{\{\tilde{\br{\Sigma}}_u\},\tilde{\br{\Sigma}}_d, \tilde{\br{\Sigma}}_{sc}\}$. The optimal solution to (SDR1.2), denoted as $\{\{\bm{W}_u\}, \bm{R}_d, \bm{R}_X\}$, can then be constructed via (\ref{eq:Wp_low_rank_qr}) and (\ref{eq:Rd_low_rank_qr}).
	
	
	After we obtain the global optimal solution to (SDR1.2), we can finally find the global optimal solution to the original problem (P1.2) via rigorously proving the tightness of the SDR and accordingly constructing the rank-one optimal beamformers, as shown in the following proposition.
	
	\begin{proposition}\label{Prop:rank_1_optimal_exist}
		\emph{Let $\{\{\bm{W}_u\}, \bm{R}_d, \bm{R}_X\}$ be an optimal solution to (SDR1.2) with $\operatorname{rank}(\bm{W}_u) \geq 1, u \in \mathcal{U}$. We could always construct a new set of solution that preserves the objective value of (SDR1.2) while still meeting all the constraints via
			\begin{align}
				\bm{\xi}_u & = \left(\bm{h}_u^H \bm{W}_u \bm{h}_u\right)^{-\frac{1}{2}} \bm{W}_u \bm{h}_u, \enspace \bar{\bm{W}}_u = \bm{\xi}_u \bm{\xi}_u^H, \enspace u \in \mathcal{U}, \\
				\bar{\bm{R}}_d & = \sum_{u=1}^{U} \bm{W}_u + \bm{R}_d - \sum_{u=1}^{U} \bar{\bm{W}}_u, 
			\end{align}
			such that $\operatorname{rank}(\bar{\bm{W}}_u) = 1.$
			\begin{proof}
				The detailed proof is similar to that for Proposition 1 and Proposition 2 in \cite{hua2023optimal} and is thus omitted for brevity.
		\end{proof}}
	\end{proposition}
	Based on Propositions \ref{Prop:Rx_span} and \ref{Prop:rank_1_optimal_exist}, we obtain the global optimal solution to (P1) via SDR with reduced complexity.

	\section{SINR-constrained Minimum Target Illumination/Echo Signal Power Maximization}\label{Sec:Max-min}
	
	Next, we consider the sensing metrics in (\ref{eq:Tx_gain}) and (\ref{eq:Rx_gain}), i.e., \textit{Metric II} and \textit{Metric III}, respectively. Accordingly, the SINR-constrained minimum target illumination power maximization problem is formulated as
	\begin{subequations}
		\begin{align}\label{eq:beampattern_sensing_RT_ori}
			(\text{P2}): & \max_{\{\bm{w}_u\}_{u=1}^U,\bm{R}_d \succeq \bm{0}, \bm{R}_X} \min_{\{\br{l}_k\}} E^t(\br{l}_k) \\
			\text{s.t. } & \text{(\ref{eq:SINR}), (\ref{eq:Sum_power}), \& (\ref{eq:Rx})},
		\end{align}
	\end{subequations}
	and the SINR-constrained minimum target echo signal power maximization problem is formulated as
	\begin{subequations}
		\begin{align}\label{eq:beampattern_sensing_RX_ori}
			(\text{P3}): & \max_{\{\bm{w}_u\}_{u=1}^U,\bm{R}_d \succeq \bm{0}, \bm{R}_X} \min_{\{\br{l}_k\}} E^r(\br{l}_k) \\
			\text{s.t. } & \text{(\ref{eq:SINR}), (\ref{eq:Sum_power}), \& (\ref{eq:Rx})}.
		\end{align}
	\end{subequations}
	
	First, we focus on solving problem (P2), as (P3) can be solved in a similar manner. By introducing the auxiliary variable 
	\begin{align}
		\mu = \min_{\{\br{l}_k\}} \bm{v}^T(\br{l}) (\sum_{u=1}^U \bm{w}_u \bm{w}_u^H +\bm{R}_d) \bm{v}^*(\br{l}),
	\end{align}
	and letting $\bm{W}_u = \bm{w}_u \bm{w}_u^H, u \in \mathcal{U}$,
	(P2) is transformed into
	\begin{subequations}
		\begin{align}\label{eq:beampattern_sensing_Rx_SDP}
			(\text{P2.1}): & \max_{\{\bm{W}_u\}_{u=1}^U,\bm{R}_d \succeq \bm{0}, \bm{R}_X} \text{ } \mu \\
			\text{s.t. } & \bm{v}^T(\br{l}) (\sum_{u=1}^U \bm{W}_u +\bm{R}_d) \bm{v}^*(\br{l}) \geq \mu, \enspace  \forall \br{l} \in \{\br{l}_k\},\\
			\label{eq:SINR_P31}
			&  \frac{\text{tr}\left(\bm{h}_u \bm{h}_u^H \bm{W}_{u}\right)}{\Gamma_u}-\sum\limits_{k \in \mathcal{U} \atop k \neq u} \text{tr}\left(\bm{h}_u \bm{h}_u^H \bm{W}_{k}\right) \\
			\nonumber
			& \quad -  \text{tr}\left(\bm{h}_u \bm{h}_u^H \bm{R}_{d}\right)-\sigma_{c}^{2} \geq 0, \enspace \forall u \in \mathcal{U},\\
			\label{eq:Sum_power_P31}
			&  \sum_{u=1}^U \operatorname{tr}(\bm{W}_u) +\operatorname{tr}(\bm{R}_d) \leq P_T, \\
			\label{eq:rank_1_P31}
			& \bm{W}_u \succeq \bm{0}, \enspace \operatorname{rank}(\bm{W}_u) = 1,  \enspace \forall u \in \mathcal{U},\\
			\label{eq:Rx_P31}
			& \bm{R}_X = \bm{R}_d + \sum_{u=1}^{U} \bm{W}_u.
		\end{align}
	\end{subequations}
	By removing the rank-one constraints, (P2.1) is relaxed as (SDR2.1), which is also an SDP. Fortunately, (SDR2.1) also enjoys a low-rank solution structure, as stated in the following.
	\begin{proposition}\label{Prop:Rx_span_beam}
		\emph{For any global optimal solution to (SDR2.1), denoted as $\{\{\bm{W}_u\},\bm{R}_d, \bm{R}_X\}$, their conjugates $\bm{W}_u^*, u \in \mathcal{U}$, and $\bm{R}_d^*$ each have a maximum rank of $K+U$ (assuming $K+U \leq N$) with eigenvectors all belonging to the subspace spanned by the columns of $\bm{V}$ and $\bm{H}_c^*$, i.e., 
			\begin{align}
				\label{eq:Wp_low_rank_beam}
				\bm{W}_u^* & = \br{U}_{be} \br{\Xi}_u \br{U}_{be}^H, \enspace u \in \mathcal{U}, \\
				\label{eq:Rd_low_rank_beam}
				\bm{R}_d^* & = \br{U}_{be} \br{\Xi}_d \br{U}_{be}^H.
			\end{align}
			As such, the optimal $\bm{R}_X$ is written as
			\begin{align}
				\bm{R}_X = \br{U}_{be}^* \br{\Xi}_{be}^* \br{U}_{be}^T,
			\end{align}
			where $\br{\Xi}_{be} = \sum_{u=1}^{U}\br{\Xi}_u + \br{\Xi}_d$ is a $(K+U) \times (K+U)$ positive semi-definite matrix, and $\br{U}_{be}$ is given as
			\begin{align}
				\label{eq:U_b_expression}
				\br{U}_{be} = \left[ \arraycolsep=0.35pt\def\arraystretch{1.2}
				\bm{H}_c^* (\bm{H}_c^T \bm{H}_c^*)^{\text{-}\frac{1}{2}}  \enspace
				\bm{V} (\bm{V}^H \bm{V})^{\text{-}\frac{1}{2}} \right].
			\end{align}
			\begin{proof}
				The proof is similar to that for Proposition \ref{Prop:Rx_span} and thus omitted.
		\end{proof}}
	\end{proposition}
	Once we obtain the optimal solution to (SDR2.1) with reduced complexity via Proposition \ref{Prop:Rx_span_beam}, it is easy to show that the SDR is tight for (P2.1),  similarly as for Proposition \ref{Prop:rank_1_optimal_exist}, i.e., there always exists a global optimal solution to (SDR2.1), denoted as $\{\{\bar{\bm{W}}_u\},\bar{\bm{R}}_d,\bar{\bm{R}}_X\}$, such that $\operatorname{rank}(\bar{\bm{W}}_u) = 1, u \in \mathcal{U}$. Therefore, problem (P2.1) or (P2) is solved. 
	
	Next, we focus on solving problem (P3). Similar to problem (P2), (P3) can be transformed into
	\begin{subequations}
		\begin{align}\label{eq:beampattern_echo_Rx_ori}
			(\text{P3.1}): & \max_{\{\bm{w}_u\}_{u=1}^U,\bm{R}_d \succeq \bm{0}, \bm{R}_X} \mu \\
			\text{s.t. } & \frac{\bm{v}^T(\br{l}) (\sum\limits_{u=1}^U \bm{W}_u +\bm{R}_d) \bm{v}^*(\br{l})}{\beta(\br{l})} \geq \mu, \enspace  \forall \br{l} \in \{\br{l}_k\},\\
			\nonumber
			& \text{(\ref{eq:SINR_P31}), (\ref{eq:Sum_power_P31}), (\ref{eq:rank_1_P31}), \& (\ref{eq:Rx_P31})},
		\end{align}
	\end{subequations}
	where $\beta(\br{l}) = 1/\left(\|\bm{a}(\br{l})\|^2 |b_{\br{l}}|\right)$. 
Since (P3.1) has a similar structure as (P2.1) except for additional weights $\{\beta(\br{l})\}$, (P3.1) can be solved via the same procedure for solving (P2.1).
It is worth noting that when there exists only one target, (P2.1) and (P3.1) correspond to the same design; while when there are multiple targets, they become different in general.


\section{Special Case with Single CU and Single Target}

In this section, we discuss the special case with one single target and one single CU, i.e., $K=1$ and $U=1$, to gain more insights.

\subsection{Optimal Communication-Sensing Tradeoffs}\label{sec:opt_cs_tradeoff}

Under the special case with \textit{Metric I}, (P1) is simplified as 
\begin{align}\label{eq:CRB_ISAC_PI_single}
	(\text{P1.s}): & \min_{\bm{w},\bm{R}_d\succeq \bm{0}, \bm{R}_X} \text{ } \text{CRB} (\bm{R}_X) \\
	\nonumber
	\text{s.t. } & \gamma (\bm{w}, \bm{R}_d) \geq \Gamma,   \|\bm{w}\|^2 +\operatorname{tr}(\bm{R}_d) \leq P_T.
\end{align}
In the following, we first find the maximum $\Gamma$ (denoted as $\Gamma_c$) that can be chosen such that (P1.s) is feasible and the maximum $\Gamma$ (denoted as $\Gamma_s$) that can be chosen such that the SINR constraint in (P1.s) is inactive. Besides, we discuss the design strategies at $\Gamma = \Gamma_s$ and $\Gamma = \Gamma_c$, respectively. With $\Gamma_c$ and $\Gamma_s$, we can efficiently swipe $\Gamma$ from $\Gamma_s$ to $\Gamma_c$, and solve the corresponding (P1.s) to characterize the complete communication-sensing tradeoff.

First, $\Gamma_c$ is easily found via maximizing the SINR under the sum power constraint. 
Let $\text{P}_c$ denote the design and it is easy to see that its optimal solution is $\bm{R}_d^c = \bm{0}$ and $\bm{w}^c = \sqrt{P_T} \frac{\bm{h}}{\|\bm{h}\|}$, i.e., maximum ratio transmission (MRT) strategy is optimal. Accordingly, we have $\Gamma_c = \frac{P_T\|\bm{h}\|^2}{\sigma_c^2}$ and the corresponding CRB is $\text{CRB}_c \triangleq \text{CRB}(P_T \frac{\bm{h}\bm{h}^H}{\|\bm{h}\|^2})$.

Next, $\Gamma_s$ is found by first solving (P1.s) without considering the SINR constraint to yield the corresponding $\bm{R}_X^s$ and then finding out the corresponding $\bm{w}^s$ and $\bm{R}_d^s$ that maximize the SINR of the CU while preserving $\bm{R}_X^s = \bm{R}_d^s + \bm{w}^s (\bm{w}^s)^H$. 
To do that, we first get the truncated EVD of $\bm{R}_X^s$ as $ \bm{R}_X^s = \br{U} \bm{\Sigma}_s \br{U}^H$, where $\br{U} \in \mathbb{C}^{N \times r_s}$ with $r_s = \operatorname{rank}(\bm{R}_X^s) \leq 4$ according to Proposition \ref{Prop:Rx_span}. Let $\text{P}_s$ denote the design at $\Gamma_s$. We then have the following Lemma.
\begin{lemma}\label{lemma:w_s_range_RXs}
	\emph{At $\text{P}_s$, $\bm{w}^s$ lies in the range space of $\bm{R}_X^s$, or equivalently $\mathcal{R}(\br{U})$.\\ 
		\begin{proof}
			See Appendix \ref{proof:ws_in_R_Xs}.
	\end{proof}}
\end{lemma}
We then maximize the SINR while preserving $\bm{R}_X^s$, i.e.,
\begin{align}
	\max_{\bm{w}^s, \bm{R}_d^s \succeq \bm{0}} \frac{|\bm{h}^H \bm{w}^s|^2}{\bm{h}^H \bm{R}_d^s \bm{h} + \sigma_c^2}, \enspace \text{s.t. } \bm{R}_X^s = \bm{R}_d^s + \bm{w}^s (\bm{w}^s)^H. 
\end{align} 
With Lemma \ref{lemma:w_s_range_RXs}, $\bm{w}^s$ is given as $\bm{w}^s = \br{U} \bm{x}$ and we thus have
\begin{align}
	\max_{\bm{x}, \bm{R}_d^s \succeq \bm{0}} |\bm{h}^H \br{U} \bm{x}|^2, \enspace \text{s.t. } \bm{R}_X^s = \bm{R}_d^s + \br{U} \bm{x} \bm{x}^H \br{U}^H.
\end{align}
It is easy to see that the optimal solution of $\bm{x}$ should admit the form of $\bm{x} = \alpha \frac{\br{U}^H \bm{h}}{\|\br{U}^H \bm{h}\|}$, where $\alpha$ is a scaling factor that has to be decided under the constraint of $\bm{R}_d^s \succeq \bm{0}$. Accordingly, $\bm{R}_d^s = \br{U}(\bm{\Sigma}_s - |\alpha|^2 \frac{\br{U}^H \bm{h} \bm{h}^H \br{U} }{\|\br{U}^H \bm{h}\|^2}) \br{U}^H$. We then gradually increase $|\alpha|^2$ from 0 to $P_T$ and find $\alpha^\star$ such that $\lambda_{\text{min}}(\bm{\Sigma}_s - |\alpha^\star|^2 \frac{\br{U}^H \bm{h} \bm{h}^H \br{U} }{\|\br{U}^H \bm{h}\|^2}) = 0$, or equivalently, $\lambda_{\text{min}}(\bm{R}_d^s) = 0$. Since $r_s \leq 4$, the seeking procedure of $\alpha^\star$ is very efficient. We then have $\bm{w}^s = \alpha^\star  \frac{\br{U} \br{U}^H \bm{h}}{\|\br{U}^H \bm{h}\|}$ and $\bm{R}_d^s = \bm{R}_X^s - \bm{w}^s (\bm{w}^s)^H$, i.e., at $\text{P}_s$, the optimal beamforming vector for the CU is achieved through projecting its channel onto the range space of $\bm{R}_X^s$, together with greedy scaling via $\alpha^\star$, and $\bm{R}_d^s$ is then added to preserve the sensing performance.
Accordingly, $\Gamma_s = \gamma(\bm{w}^s, \bm{R}_d^s)$ and the corresponding CRB is $ \text{CRB}_s \triangleq \text{CRB}(\bm{R}_X^s)$.


It is also worth discussing the CRB and the SINR achieved under the design with \textit{Metric II} or \textit{Metric III}  ((P2) or (P3)) without considering communication requirements. This is denoted as design $\text{P}_{s'}$. At $\text{P}_{s'}$, we have $\bm{R}_{X}^{s'} = P_T \frac{\bm{v}^* \bm{v}^T}{\|\bm{v}\|^2}$ with $\bm{w}^{s'} = \sqrt{P_T} \frac{\bm{v}^*}{\|\bm{v}\|}$, i.e., $\bm{w}^{s'}$ is used for both sensing and communication. The corresponding CRB and SINR are $\text{CRB}_{s'} \triangleq \text{CRB}(P_T \frac{\bm{v}^*\bm{v}^T}{\|\bm{v}\|^2})$ and $\Gamma_{s'} \triangleq \frac{P_T |\bm{h}^H \bm{v}^*|^2}{\|\bm{v}\|^2 \sigma_c^2}$, respectively. 

\subsection{Analysis with Collocated Target and CU} 

We further consider the scenario where the target is collocated with the CU and there is only LoS path between the Tx and the target/CU, i.e., $\bm{h} = \bm{v}^*$. Under the far-field scenario, it has been revealed in \cite{hua2022mimojournal} that regardless of the target angle, the optimal transmission strategy is the same for both sensing in terms of target angle CRB minimization and communication in terms of SINR or achievable rate maximization, i.e., $\text{P}_c$ is same as $\text{P}_s$. To achieve that, we simply steer the beam towards the angle where the target/CU is located \cite{li2007range}. However, under the considered 3D near-field scenario, this intuitive result needs a closer scrutiny. Indeed, under the case of collocated target/CU with $\bm{h} = \bm{v}^*$, it follows from Section \ref{sec:opt_cs_tradeoff} that the optimal $\bm{R}_X$ at $\text{P}_c$ is same as that at $\text{P}_{s'}$, i.e., $\bm{R}_X^c = \bm{R}_X^{s'} = P_T \frac{\bm{v}^*\bm{v}^T}{\|\bm{v}\|^2}$,  while based on Proposition \ref{Prop:Rx_span}, $\bm{R}_X^s$ at $\text{P}_s$ is generally expressed as $\bm{R}_X^s = \br{U}_{r}^* \br{\Sigma}_{r}^* \br{U}_{r}^T$, with $\br{U}_{r} = \left[\frac{\bm{v}}{\|\bm{v}\|}, \frac{\dot{\bm{v}}_\mr{x}}{\|\dot{\bm{v}}_\mr{x}\|},\frac{\dot{\bm{v}}_\mr{y}}{\|\dot{\bm{v}}_\mr{y}\|},\frac{\dot{\bm{v}}_\mr{z}}{\|\dot{\bm{v}}_\mr{z}\|}\right]$. 
An interesting relevant question is thus whether $\br{\Sigma}_{r} = \operatorname{diag}(P_T,0,0,0)$ or $\bm{R}_X^s = \bm{R}_X^c$ also holds under the 3D near-field scenario.
In the following, we identify the special configurations where 
$\bm{R}_X^s = \bm{R}_X^c = P_T \frac{\bm{v}^*\bm{v}^T}{\|\bm{v}\|^2}$.

\begin{figure}[t]
	\centering
	\includegraphics[width=3.5in]{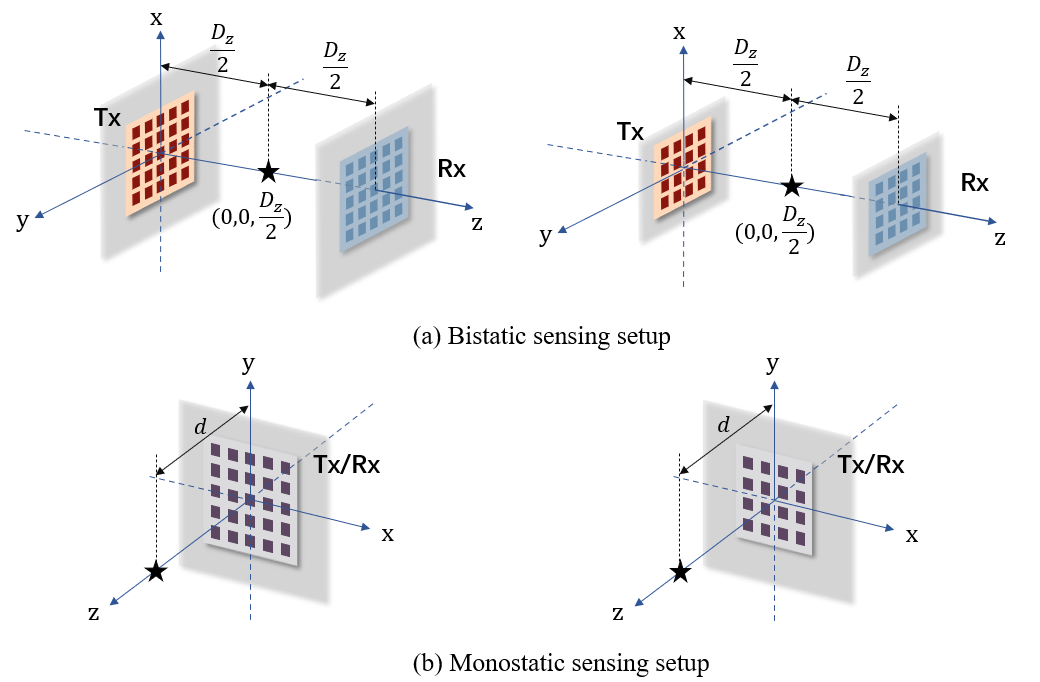} 
	\centering
	\caption{Two special configurations with the collocated target/CU and UPA with $M=N=n^2$ where $\bm{R}_X^s = \bm{R}_X^c$: (a). A bistatic sensing setup with the collocated target/CU located in the middle of the symmetric Tx and Rx; and (b). A monostatic sensing setup with the position of the target/CU being $(0,0,d), d>0$ and symmetric Tx/Rx. The collocated target/CU is denoted by the black star. The number of antennas can be odd or even, as long as the Tx/Rx array is centered at the $\mr{z}$-axis.}
	\label{fig:CRB_d0_single_setup_isac}
\end{figure}


\begin{proposition}\label{prop:CRB_min_MRC}
	\emph{Under the configurations in Fig. \ref{fig:CRB_d0_single_setup_isac}, where the target/CU is present towards the middle of a symmetric UPA, it follows that $\bm{R}_X^s = \bm{R}_X^c = \frac{P_T}{\|\bm{v}\|^2} \bm{v}^* \bm{v}^T$.
		\begin{proof}
			See Appendix \ref{proof:single_special_merged_Pc_Ps}.
	\end{proof}}
\end{proposition}


However, under other more general configurations, e.g., the target/CU is located elsewhere or the UPA is not symmetric, we have $\bm{R}_X^c \neq \bm{R}_X^s$, i.e., 
the transmission strategy for sensing (i.e., CRB minimization) and that for communication (i.e., SINR maximization) are different,
as will be shown in Section \ref{sec:numerical} by simulations. This is in sharp contrast to that in the far-field scenario, where $\bm{R}_X^s = \bm{R}_X^c$ always holds regardless of the collocated target/CU location.

\section{Numerical results}\label{sec:numerical}

This section evaluates the three designs proposed in Section \ref{sec:CRB_min} and
Section \ref{Sec:Max-min}.



\subsection{Special Case with Single Collocated Target and CU}\label{sec:sepcial_colocated}

In this subsection, we first consider the case of collocated target/CU.
Since the results are similar, we focus on the bistatic sensing setup in Fig. \ref{fig:CRB_d0_single_setup_isac}(a), where the target/CU can move along the y-axis with position $(0,d,D_z/2)$ by varying $d$. Its reflection coefficient is assumed to be $b = 1$. We consider LoS path between the Tx array and the target/CU, i.e., $\bm{v} = \bm{h}^*$. The spacing between adjacent antennas is half-wavelength with $f_c = 28$ GHz. The UPA at both Tx and Rx has a rectangular shape with $M = N = n_\mr{x} \times n_\mr{y} = n^2 =  48 \times 48 = 2304$, where $n_\mr{x}$ and $n_\mr{y}$ denote the numbers of antennas along the $\mr{x}$-axis and the $\mr{y}$-axis, respectively. The total transmit power budget is assumed to be $P_T = $10 dBm and $\bm{Q} = \sigma_s^2 \bm{I}$ with $\sigma_s^2 = -50$ dBm. Besides, $\sigma_c^2 = -50$ dBm. As a result, the threshold value to define the near-field region is computed as $d_\text{nf} = 2 D^2/\lambda = 24.7$ m with $D = \sqrt{(n_x^2+n_y^2)s^2} = 0.364$ m \cite{cong2023near}. We further set $D_z = d_\text{nf}/10 = 2.47$ m. 
For the ease of graphical illustration, we show in the following the corresponding 3D beampattern projected onto the 2D $\mr{y}-\mr{z}$ plane, which is defined as
\begin{align}\label{eq:2D_beampattern}
	\mathbb{P}(0,\mathrm{y},\mathrm{z}) = \bm{v}^T(0,\mathrm{y},\mathrm{z}) \bm{R}_X \bm{v}^*(0,\mathrm{y},\mathrm{z}).
\end{align}

\begin{figure*}[t]
	\centering
	\setlength{\abovecaptionskip}{+4mm}
	\setlength{\belowcaptionskip}{+1mm}
	\subfigure[Beampattern with $d = 0$.]{ \label{fig:Beampatter_unify_d0_bi}
		\includegraphics[width=1.8in]{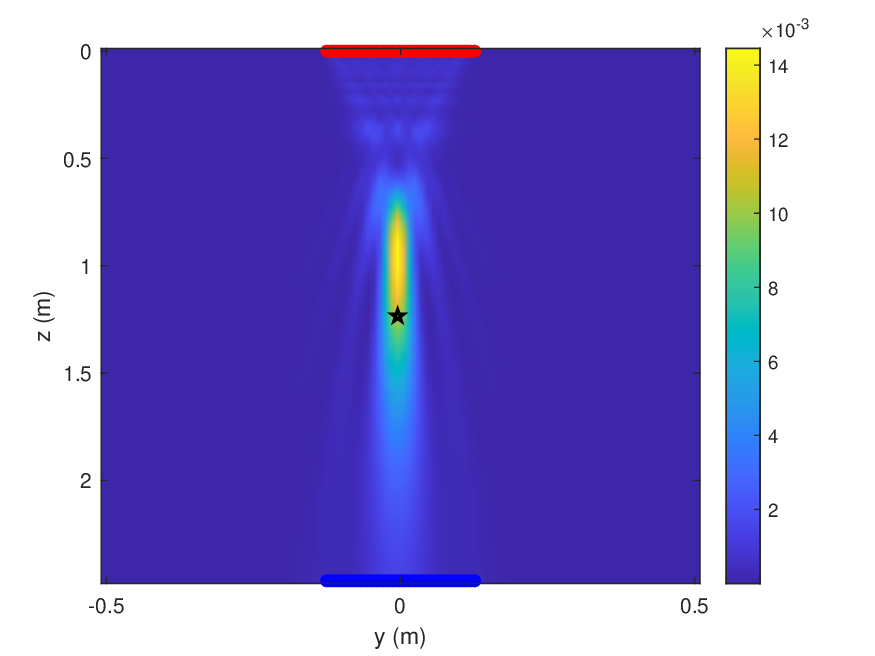}}
	\subfigure[CRB versus $d$.]{ \label{fig:CRB_d_swipe_single_target}
		\includegraphics[width=1.8in]{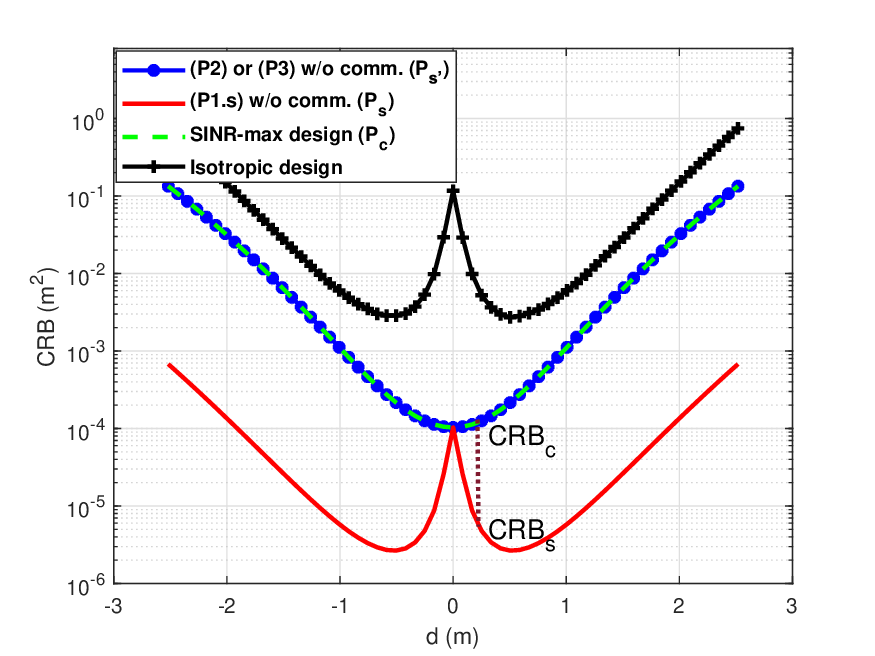}}
	\subfigure[SINR versus $d$.]{ \label{fig:Rate_d_swipe_single_target}
		\includegraphics[width=1.8in]{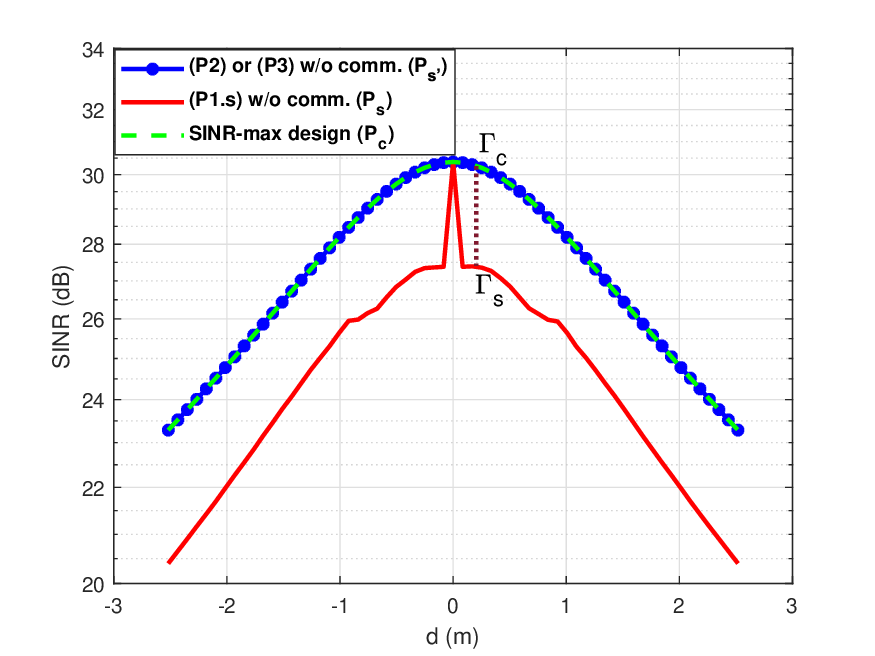}}
	\subfigure[Beampattern via $\text{P}_s$.]{ \label{fig:single_beam_CRB}
		\includegraphics[width=1.7in]{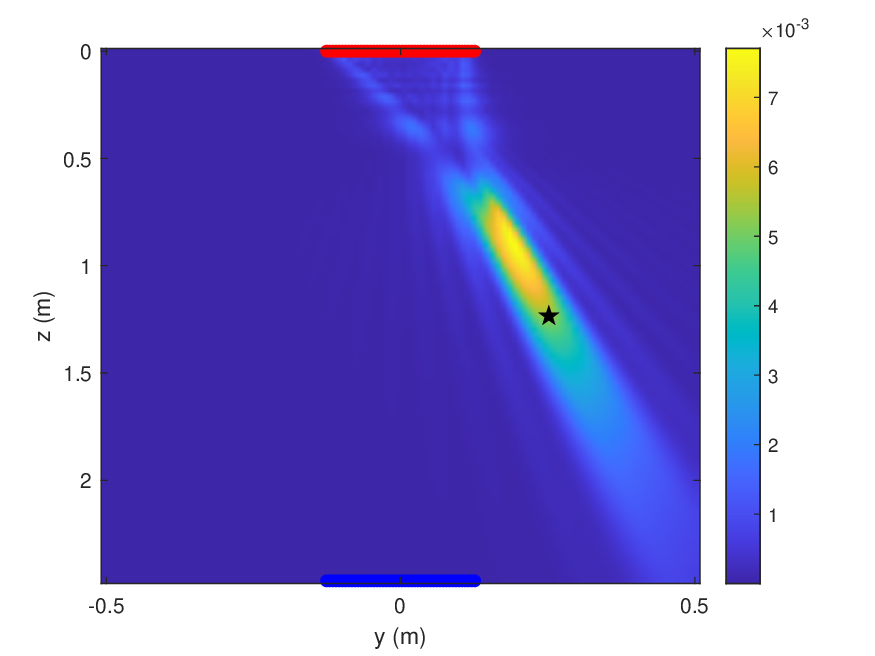}}
	\subfigure[Beampattern via $\text{P}_c$ and $\text{P}_{s'}$.]{ \label{fig:single_beam_rate_max}
		\includegraphics[width=1.7in]{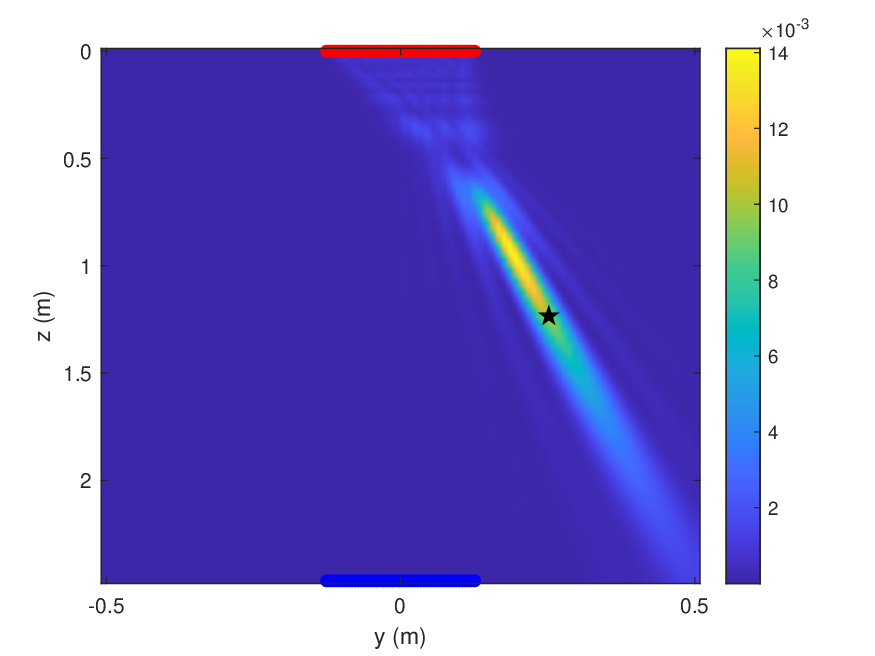}}
	\subfigure[Comparison of eigenvalue of $\bm{R}_X$.]{ \label{fig:single_eigen_comp}
		\includegraphics[width=1.7in]{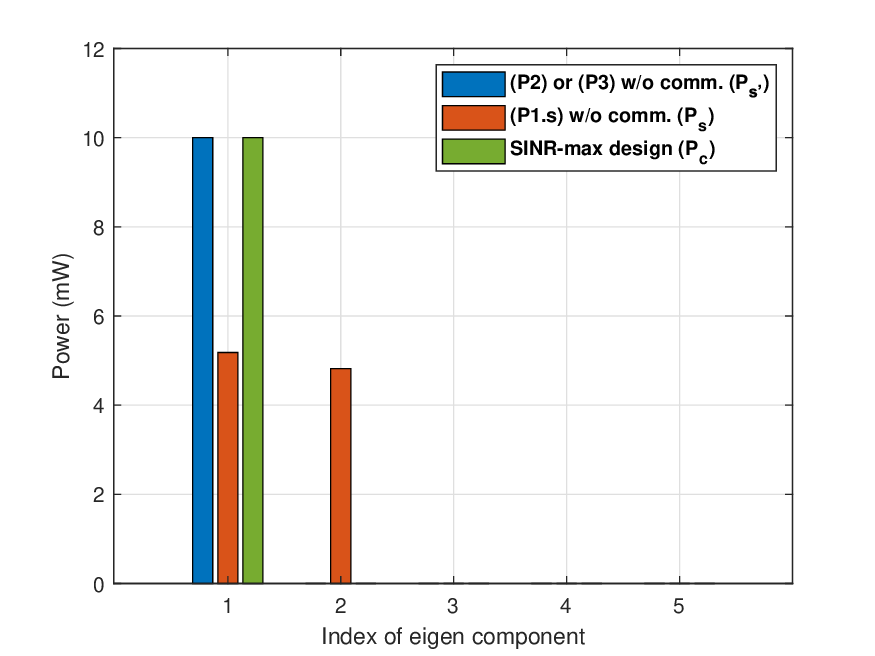}}
	\subfigure[The tradeoff curve.]{ \label{fig:single_merged_CR_region}
		\includegraphics[width=1.7in]{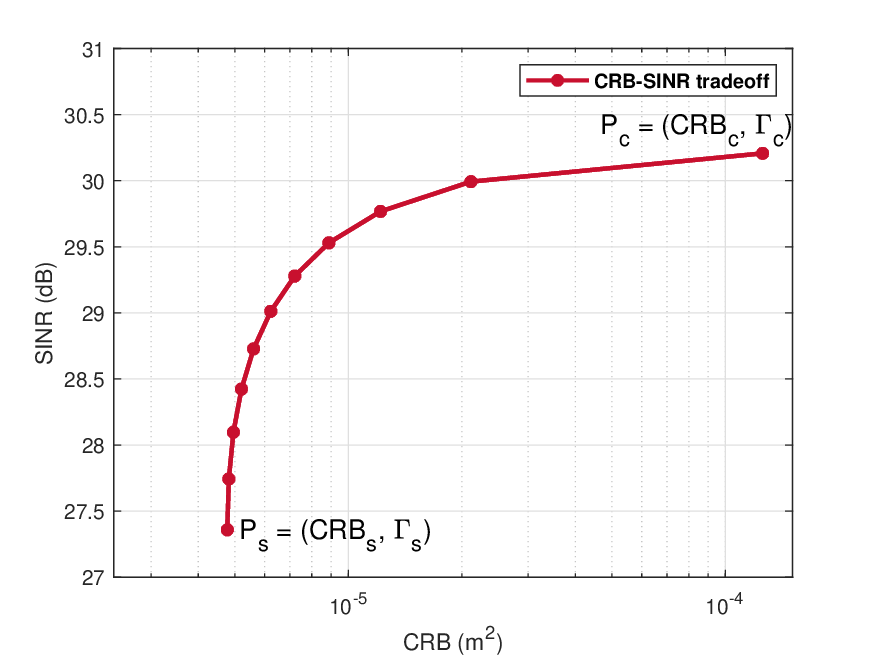}}
	\caption{The considered bistatic sensing scenario with collocated target/CU: (a). The 3D beampattern projected onto $\mr{y}-\mr{z}$ plane of three different designs with $d=0$, where the black star denotes the collocated target/CU; (b)-(c). CRB and SINR versus $d$; (d)-(e). The 3D beampattern of three designs with $d = 23.5 \lambda = 0.2518$ m; (f). The comparison of eigenvalues of $\bm{R}_X$ in three designs; and (g). The communication-sensing tradeoff curve.}
	\label{fig:single_target_comp}
\end{figure*}

We first show the resultant beampattern when $d=0$ in Fig. \ref{fig:Beampatter_unify_d0_bi}, in which (P2) or (P3) without communication requirement ($\text{P}_{s'}$), (P1.s) without communication requirement ($\text{P}_s$), and SINR-maximization design ($\text{P}_c$) all yield the same result with $\bm{R}_X = \frac{P_T}{\|\bm{v}\|^2} \bm{v}^* \bm{v}^T$. In other words, there is no tradeoff between sensing and communication in this specific configuration, which is in accordance with Proposition \ref{prop:CRB_min_MRC}. 

Next, when the collocated target/CU is moving along the y-axis with $d$ ranging from $-10 (D/\sqrt{2})$ to $10 (D/\sqrt{2})$, Fig. \ref{fig:CRB_d_swipe_single_target} and Fig. \ref{fig:Rate_d_swipe_single_target} show the corresponding CRB of the target/CU and the SINR for the CU with various designs, respectively. It is observed that the curves of CRB and SINR are symmetric with respect to the axis $d=0$. This is reasonable since both the Tx and Rx arrays are symmetrically centered at the z-axis. Besides, $\text{P}_{s'}$ is same as $\text{P}_c$, as the target is collocated with the CU.
It is also observed that when the target/CU is located at $(0,0,D_z/2)$, the optimized CRB under $\text{P}_s$ is actually the highest in that small region around $|d| \geq 0$ even now the target/CU is closest to the centers of both Tx and Rx array (with distances both equal to $D_z/2$), and when the target/CU is moving slightly away from the z-axis, the optimized CRB decreases dramatically, as sharper differences in phases among different antennas facilitate the target localization. However, when the target is moving further away from the z-axis, the CRB tends to increase again due to the decreasing amplitudes of the received signal. Similar phenomenon is also observed with isotropic transmission, i.e., $\bm{R}_X = \frac{P_T}{N}\bm{I}$. This phenomenon is quite different from the conventional wisdom in the far-field scenario, in which the CRB tends to increase monotonically when the target moves away from the transmitter/receiver. 

Besides, under other cases with $d >0$, e.g., $d = 23.5 \lambda$, it is observed from Fig. \ref{fig:CRB_d_swipe_single_target} and Fig. \ref{fig:Rate_d_swipe_single_target} that the optimal transmission strategy for communication ($\text{P}_c$) is different from that for sensing ($\text{P}_s$).
Indeed, as seen from Fig. \ref{fig:single_beam_CRB}-Fig. \ref{fig:single_beam_rate_max}, although the beampatterns obtained via these three designs all focus towards the target/CU, the one obtained from $\text{P}_s$ in Fig. \ref{fig:single_beam_CRB} is thicker, which is significantly different from those obtained from $\text{P}_{s'}$ and $\text{P}_c$ in Fig. \ref{fig:single_beam_rate_max}. While the optimal solution for $\text{P}_{s'}$ and $\text{P}_c$ is still $\bm{R}_X = \frac{P_T}{\|\bm{v}\|^2} \bm{v}^* \bm{v}^T$, the optimal solution for $\text{P}_s$ is in general of high rank, as shown in Fig. \ref{fig:single_eigen_comp}. Under the setup with $d = 23.5 \lambda$, we also show the communication-sensing tradeoff in Fig. \ref{fig:single_merged_CR_region}. Notice that the corresponding CRB values $\text{CRB}_c$, $\text{CRB}_s$ and the SINR values $\Gamma_c, \Gamma_s$ in Fig. \ref{fig:single_merged_CR_region} can be readily obtained by drawing vertical lines at $d = 23.5 \lambda$ that intersect with the $\text{P}_s$ design curve and the $\text{P}_c$ design curve in Fig. \ref{fig:CRB_d_swipe_single_target} and Fig. \ref{fig:Rate_d_swipe_single_target}, respectively.  

Based on the above results, one could conclude that for the collocated target/CU case, the optimal transmission strategies for sensing and communication are typically different under the considered 3D near-field scenario, except for those special cases shown in Fig. \ref{fig:CRB_d0_single_setup_isac}. This is in sharp contrast to that in the far-field scenario, in which the optimal transmission strategies are the same for both sensing and communication.

\begin{figure*}[t]
		\centering
		\setlength{\abovecaptionskip}{+4mm}
		\setlength{\belowcaptionskip}{+1mm}
		\subfigure[Beampattern via SINR-max ($\text{P}_c$).]{ \label{fig:separated_single_beam_rate_max}
			\includegraphics[width=1.7in]{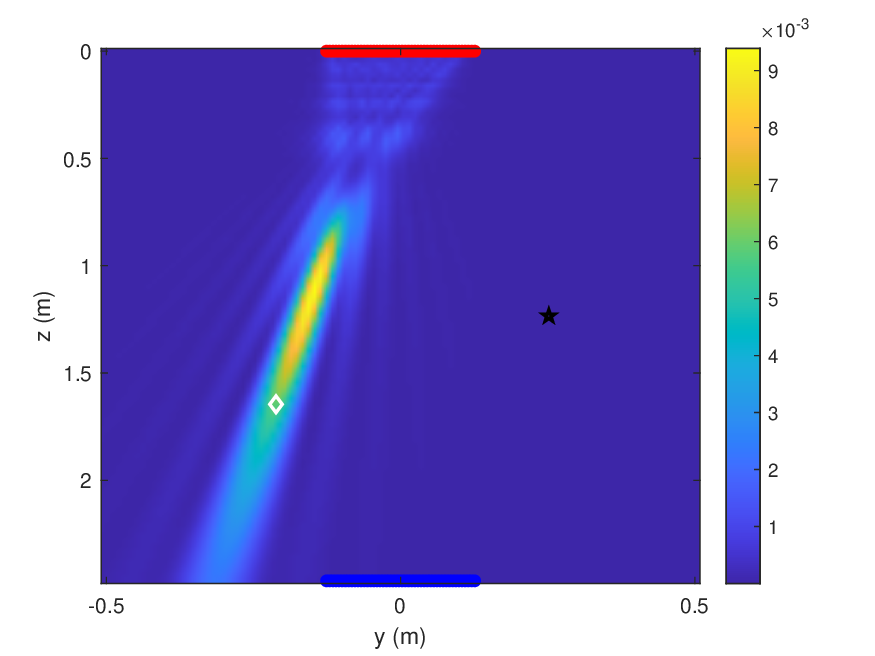}}
		\subfigure[Beampattern via CRB-min at (i).]{ \label{fig:separated_single_beam_CRB_between}
			\includegraphics[width=1.7in]{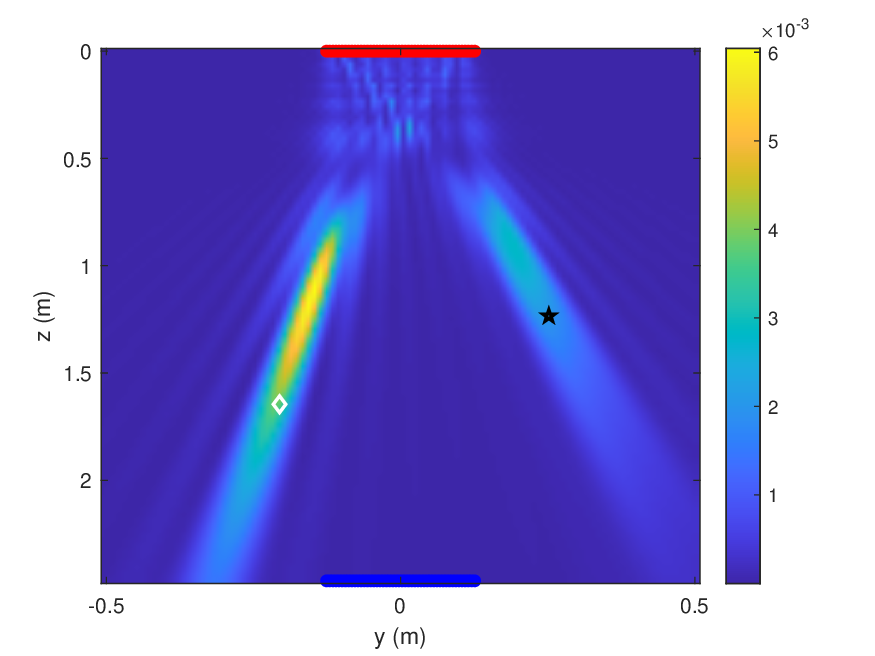}}
		\subfigure[Beampattern via Max-min at (ii).]{ \label{fig:separated_single_beam_Max_min_between}
			\includegraphics[width=1.7in]{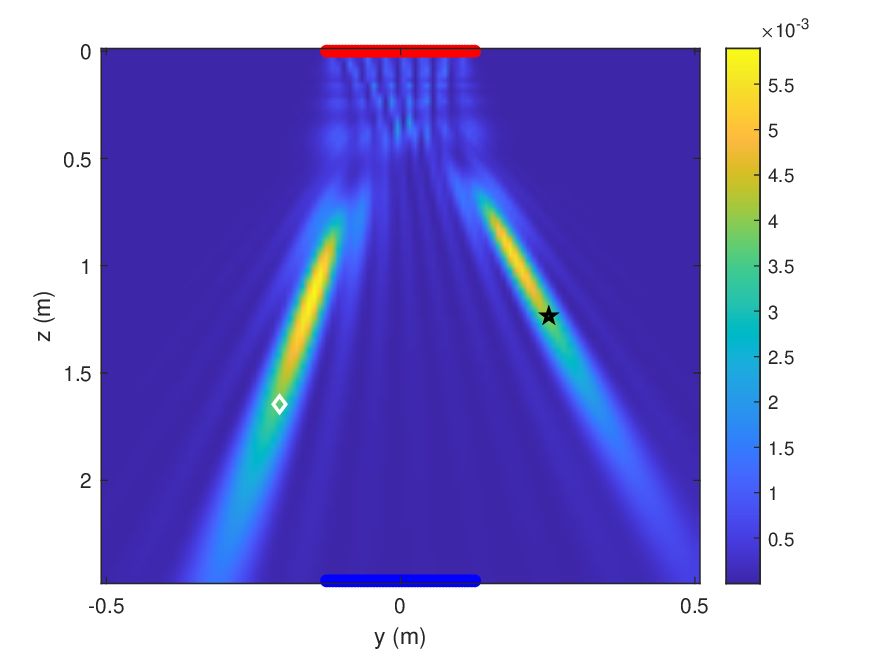}}
		\subfigure[The corresponding tradeoff.]{ \label{fig:separated_CR_region}
			\includegraphics[width=1.7in]{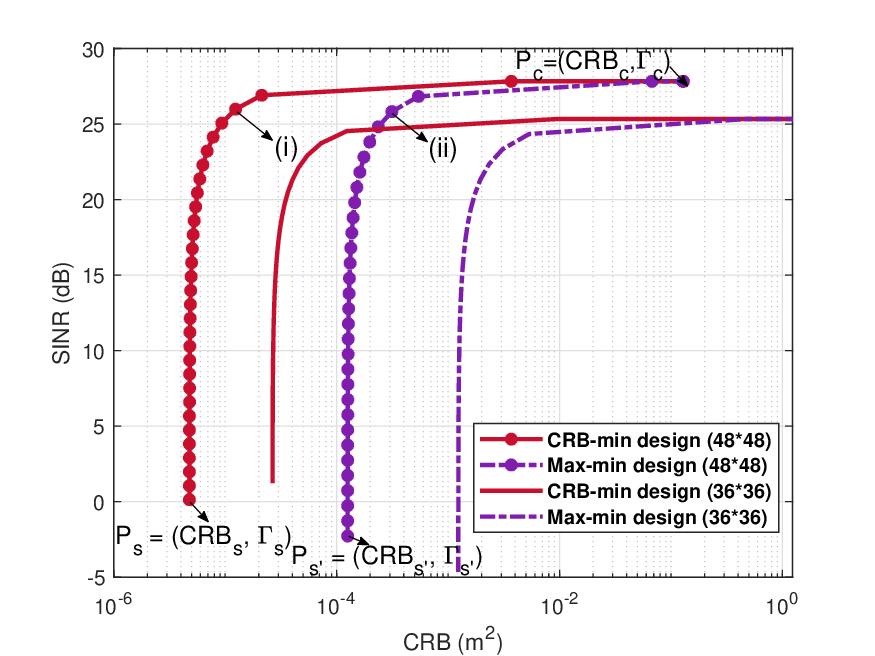}}
		\caption{(a)-(c). The comparison of the beampatterns via different designs with $\eta = 0$. The  black star denotes the target while the white diamond denotes the CU, respectively. The red line represents the Tx array looking from the top while the blue line represents the Rx array. (d). The corresponding tradeoff, in which the points corresponding to the beampatterns in (a)-(c) are also marked in the curves. 
		}
		\label{fig:isac_separate_target_CU}
\end{figure*}

\subsection{Special Case with Separated Single Target and Single CU}

Next, we consider the case of separated target and CU. In this special case, we adopt the following modeling of the communication channel between Tx and the CU,
\begin{align}\label{eq:h_comp}
	\bm{h} = \bm{h}_{\text{LoS}} + \eta \bm{h}_{\text{NLoS}},
\end{align}
where $\bm{h}_{\text{LoS}}$ is the LoS path from the Tx to the CU and $\bm{h}_{\text{NLoS}}$ is the NLoS path from the Tx to the target and then to the CU, $\eta$ is the coefficient depending on both the path-loss and radar cross section (RCS) of the target. In this subsection, we adopt the same setup as Section \ref{sec:sepcial_colocated}, except that the location of the CU is set to be $(0,-D_z/12,2D_z/3)$. Accordingly, $\bm{h}_{\text{LoS}} = \bm{v}^*_{c}$ and $\bm{h}_{\text{NLoS}} = \bm{v}^*_{t}$, where $\bm{v}_{c}$ and $\bm{v}_t$ are the steering vectors at the Tx corresponding to the CU and the target, respectively.

We first consider the case where there is only LoS path, i.e., $\eta = 0$. In Fig. \ref{fig:isac_separate_target_CU}, we show the CRB-SINR tradeoffs and the corresponding beampatterns at different points with both the design via solving (P1), denoted as CRB-min, and the design via solving (P2) or (P3), denoted as Max-min.
Since the target location remains the same as Section \ref{sec:sepcial_colocated}, the beampatterns via $\text{P}_{s}$ and $\text{P}_{s'}$ are the same as already being shown in Fig. \ref{fig:single_beam_CRB} and Fig. \ref{fig:single_beam_rate_max}, respectively. 
From Fig. \ref{fig:separated_CR_region}, it is observed that the CRB at $\text{P}_s$ is much smaller than that at $\text{P}_{s'}$ and the Max-min tradeoff curve in purple is embedded in the CRB-min design tradeoff curve in red, which shows that maximizing the illumination power or sensing beampattern gain towards the target does not lead to the optimal sensing performance in terms of positioning CRB minimization in the considered 3D near-field scenario. This is also quite different from that under the far-field scenario, where maximizing sensing beampattern gain towards the target angle yields the same solution as that obtained via minimizing the CRB for target angle estimation \cite{hua2022mimojournal}. 
Besides, it is shown that with much larger number of antennas ($M=N = 48 \times 48$ versus $36 \times 36$ at both Tx and Rx), the corresponding tradeoff curve moves towards the upper left side of the figure, which shows the benefit of ELAA and the waveform optimization in enhancing the near-field ISAC performance. 
Furthermore, beam focusing capability is demonstrated in  Fig. \ref{fig:separated_single_beam_rate_max} to Fig. \ref{fig:separated_single_beam_Max_min_between}. Specifically, from Fig. \ref{fig:separated_CR_region}, by leveraging the beam focusing capability provided by ELAA, we obtain better performance in both sensing and communication, 
such as the point marked as (i) in Fig. \ref{fig:separated_CR_region}.
\begin{figure*}[t]
		\centering
		\setlength{\abovecaptionskip}{+4mm}
		\setlength{\belowcaptionskip}{+1mm}
		\subfigure[Beampattern via SINR-max ($\text{P}_c$).]{ \label{fig:separated_single_beam_rate_max_scatter}
			\includegraphics[width=1.7in]{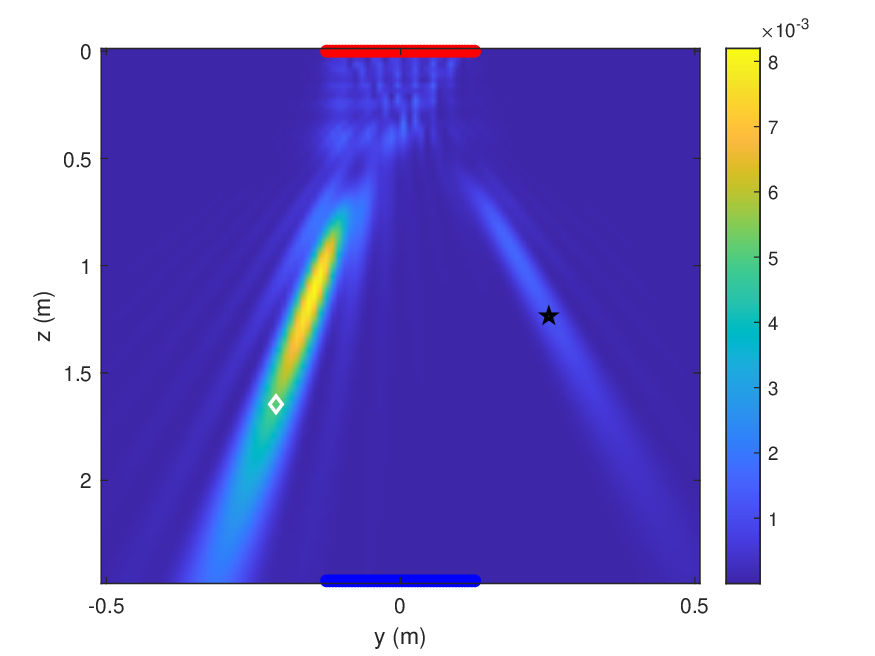}}
		\subfigure[Beampattern via CRB-min at (i).]{ \label{fig:separated_single_beam_CRB_between_scatter}
			\includegraphics[width=1.7in]{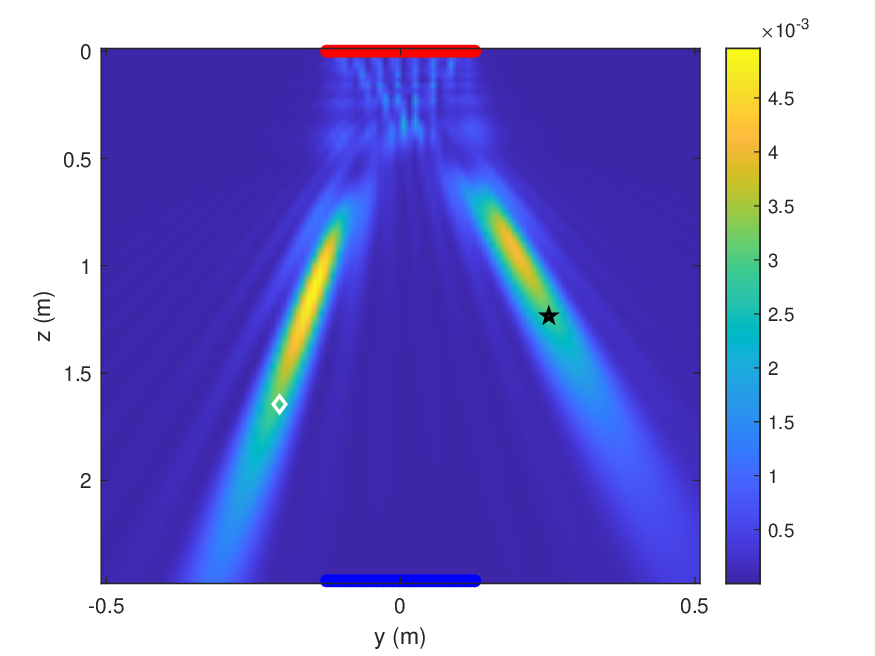}}
		\subfigure[Beampattern via Max-min at (ii).]{ \label{fig:separated_single_beam_Max_min_between_scatter}
			\includegraphics[width=1.7in]{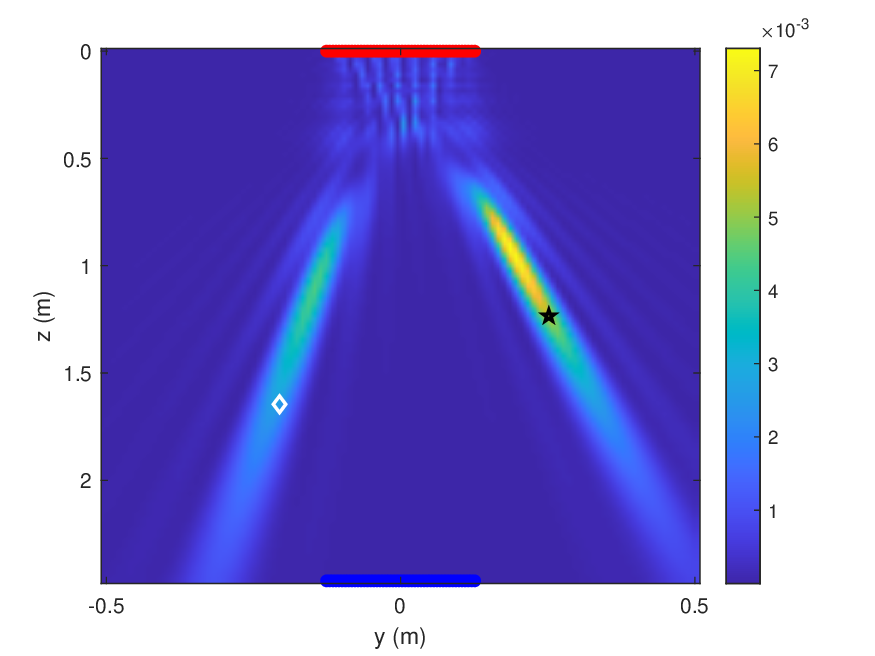}}
		\subfigure[The corresponding tradeoff.]{ \label{fig:separated_CR_region_scatter}
			\includegraphics[width=1.7in]{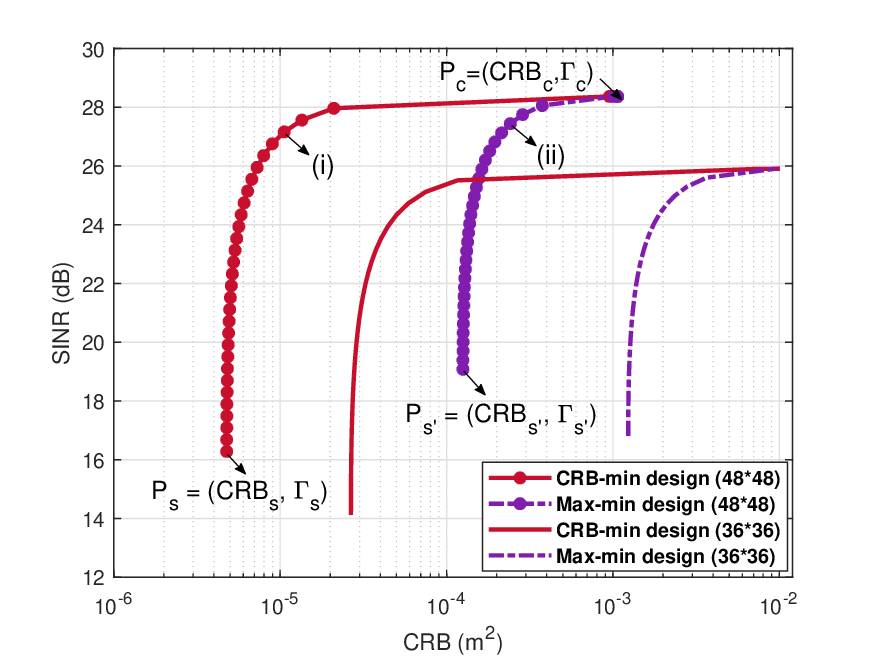}}
		\caption{(a)-(c). The comparison of the beampatterns via different designs with $\eta = 0.3$. The black star denotes the target while the white diamond denotes the CU, respectively. The red line represents the Tx array looking from the top while the blue line represents the Rx array. (d). The corresponding tradeoffs, in which the points corresponding to the beampatterns in (a)-(c) are also marked in the curves.}
		\label{fig:isac_separate_target_CU_scatter}
\end{figure*} 

We then consider the NLoS case with $\eta = 0.3$ in Fig. \ref{fig:isac_separate_target_CU_scatter}. 
Compared to the LoS case, it is observed in Fig. \ref{fig:separated_CR_region_scatter} that at $\text{P}_{s}$, the resultant SINR $\Gamma_s$ is significantly higher due to the additional NLoS path from the target to the CU. In other words, with the objective of minimizing CRB, the signal strength of the scattered path from the target to the CU also increases, leading to higher SINR. It is also observed in Fig. \ref{fig:separated_CR_region_scatter} that the CRB at $\text{P}_c$ is significantly lower than that in the LoS case, as the communication channel $\bm{h}$ in (\ref{eq:h_comp}) is composed of two components and SINR-maximization design allocates power towards both the CU and the target, as shown in Fig. \ref{fig:separated_single_beam_rate_max_scatter}. This helps reduce the CRB for localizing the target. With an additional NLoS path, we can achieve even better performance in both sensing and communication, as shown in (i) marked in Fig. \ref{fig:separated_CR_region_scatter}.
Other observations are similar to the LoS case, e.g., larger number of antennas helps enhance both sensing and communication performance, and beam focusing capability is
demonstrated in Fig. \ref{fig:separated_single_beam_rate_max_scatter}, Fig. \ref{fig:separated_single_beam_CRB_between_scatter}, and Fig. \ref{fig:separated_single_beam_Max_min_between_scatter}.


\subsection{General Case with Multiple Targets and Multiple CUs}

\begin{figure*}[t]
		\centering
		\setlength{\abovecaptionskip}{+4mm}
		\setlength{\belowcaptionskip}{+1mm}
		\subfigure[Beampattern via CRB-min with $\Gamma = 5$ dB.]{ \label{fig:2Dbeam_CRB_isac_5db}
			\includegraphics[width=1.7in]{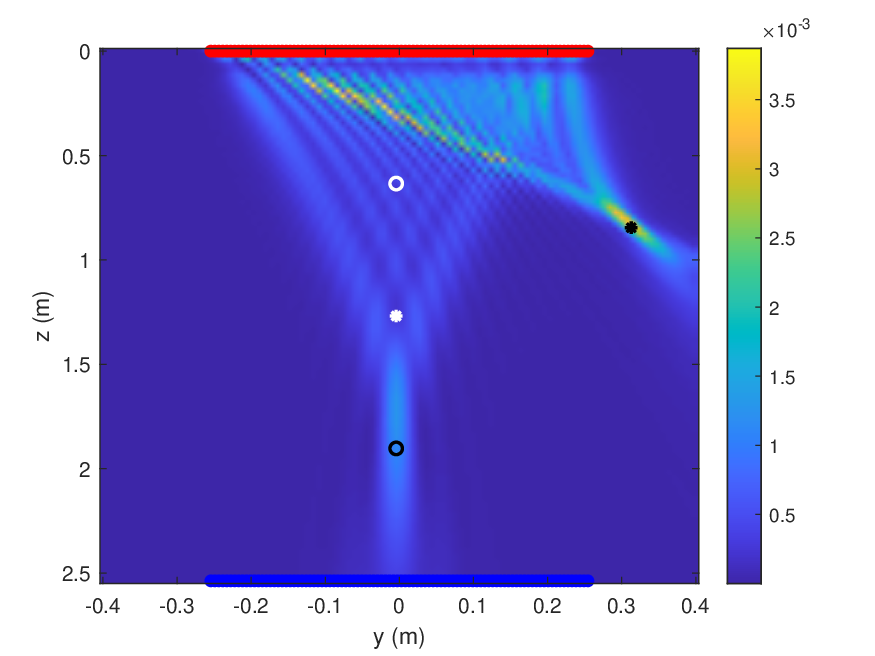}}
		\subfigure[Beampattern via Max-min Tx with $\Gamma = 5$ dB.]{ \label{fig:2Dbeam_Tx_max_min_isac_5db}
			\includegraphics[width=1.7in]{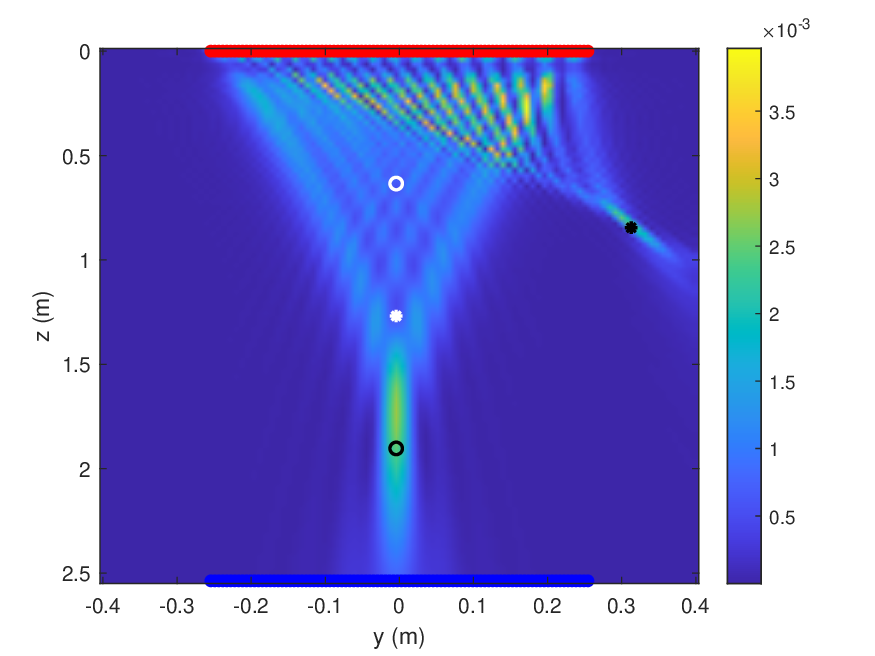}}
		\subfigure[Beampattern via Max-min Rx with $\Gamma = 5$ dB.]{ \label{fig:2Dbeam_Rx_max_min_isac_5db}
			\includegraphics[width=1.7in]{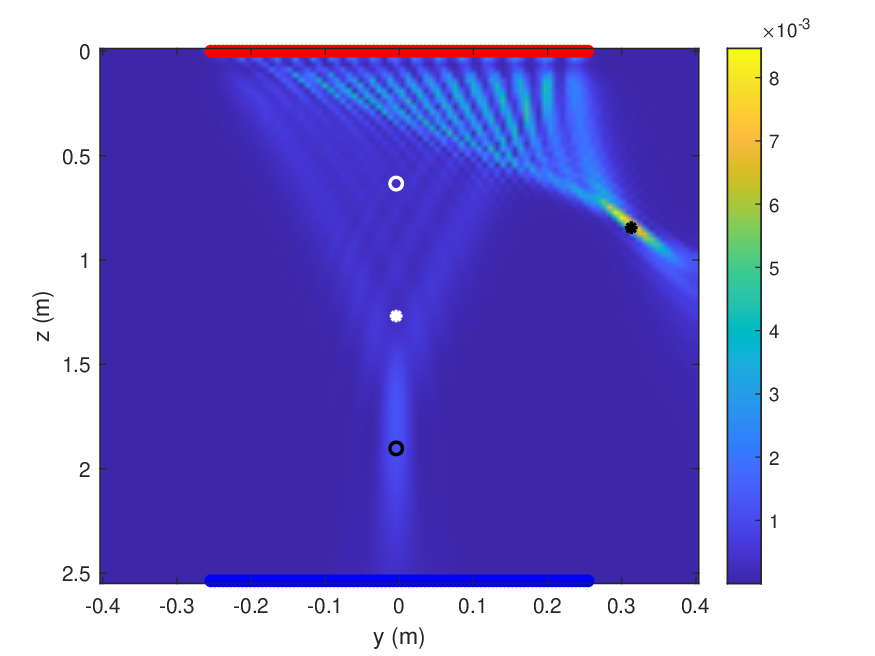}}
		\subfigure[Sum-CRB versus $\Gamma$.]{ \label{fig:isac_CRB_gamma}
			\includegraphics[width=1.7in]{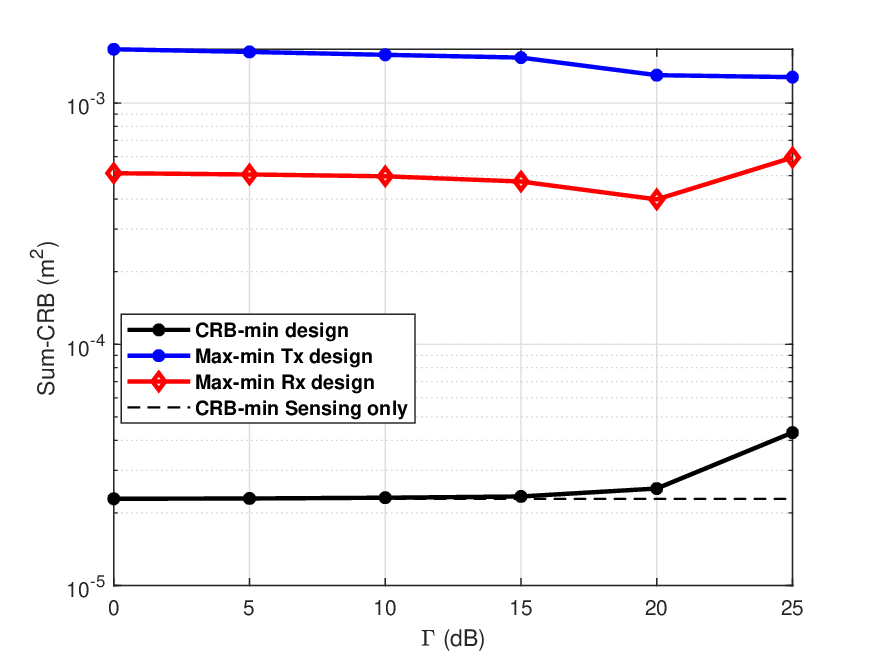}}
		\subfigure[Beampattern via CRB-min with $\Gamma = 25$ dB.]{ \label{fig:2Dbeam_CRB_isac_25db}
			\includegraphics[width=1.7in]{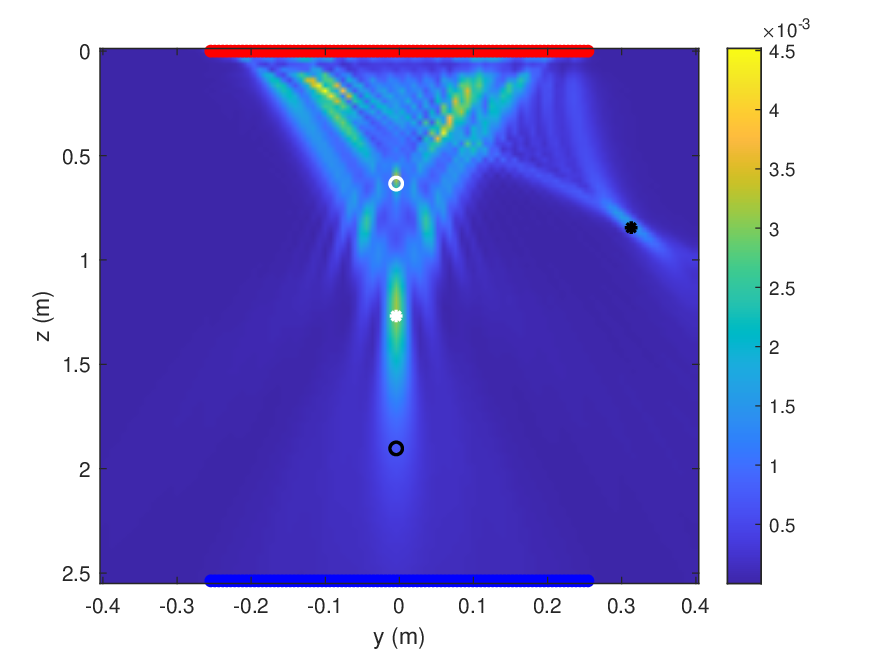}}
		\subfigure[Beampattern via Max-min Tx with $\Gamma = 25$ dB.]{ \label{fig:2Dbeam_Tx_max_min_isac_25db}
			\includegraphics[width=1.7in]{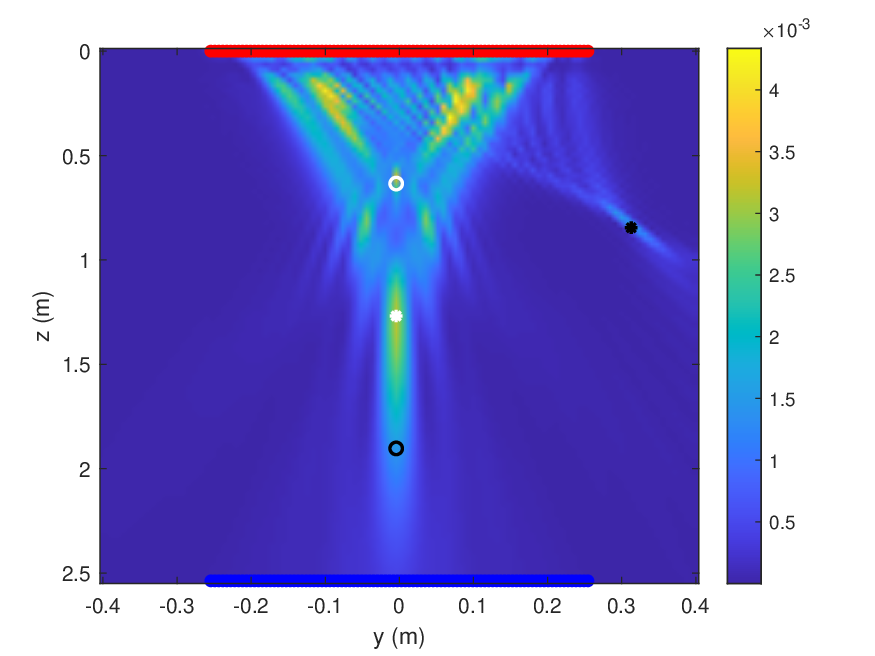}}
		\subfigure[Beampattern via Max-min Rx with $\Gamma = 25$ dB.]{ \label{fig:2Dbeam_Rx_max_min_isac_25db}
			\includegraphics[width=1.7in]{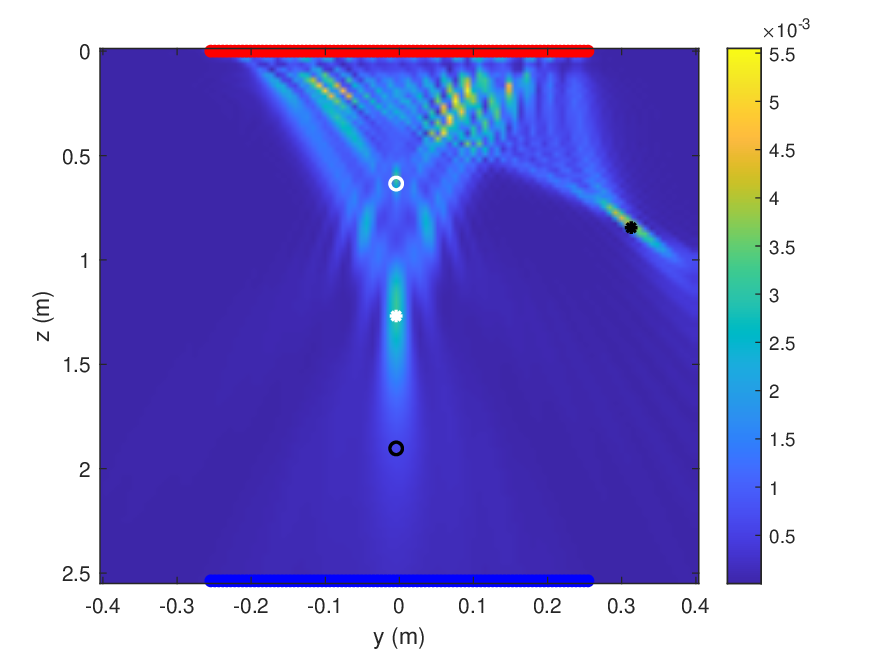}}
		\subfigure[minimum target illumination power or echo signal power versus $\Gamma$.]{ \label{fig:isac_mu_gamma}
			\includegraphics[width=1.7in]{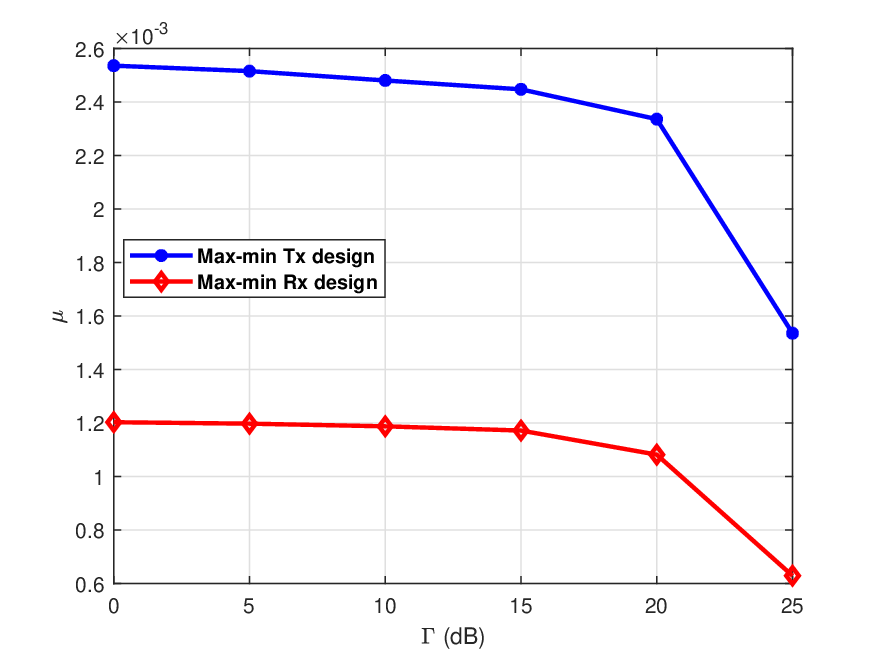}} 
		\subfigure[Beampattern $\bm{W}_1$ via CRB-min with $\Gamma = 25$ dB.]{ \label{fig:2Dbeam_CRB_W1}
			\includegraphics[width=1.7in]{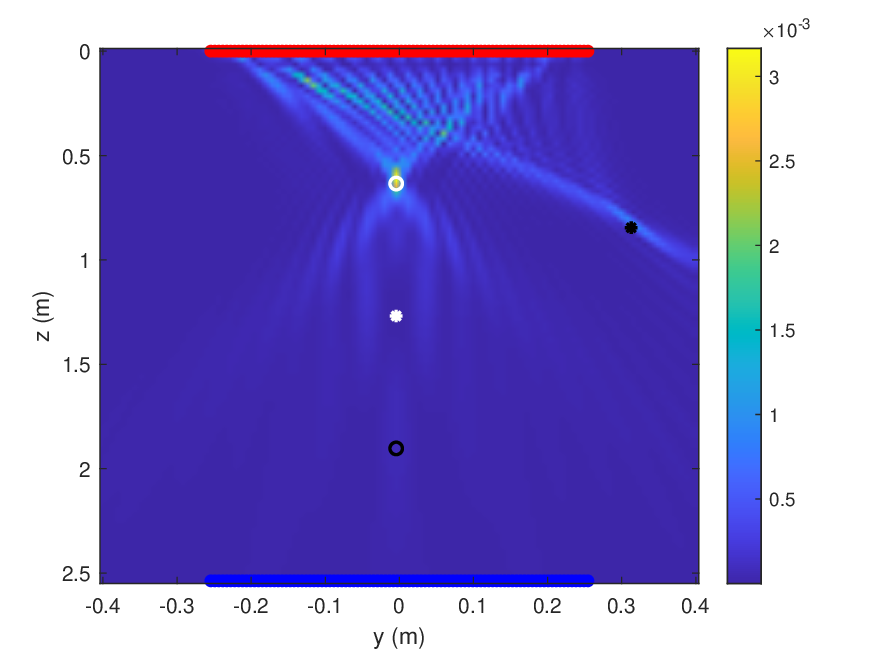}}
		\subfigure[Beampattern $\bm{W}_2$ via CRB-min with $\Gamma = 25$ dB.]{ \label{fig:2Dbeam_CRB_W2}
			\includegraphics[width=1.7in]{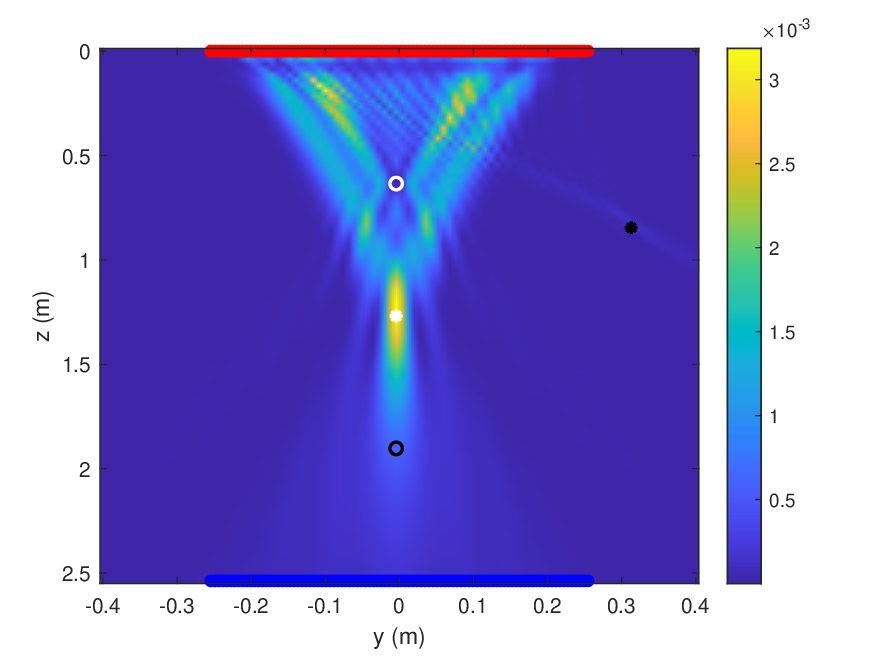}}
		\subfigure[Beampattern $\bm{R}_d$ via CRB-min with $\Gamma = 25$ dB.]{ \label{fig:2Dbeam_CRB_Rd}
			\includegraphics[width=1.7in]{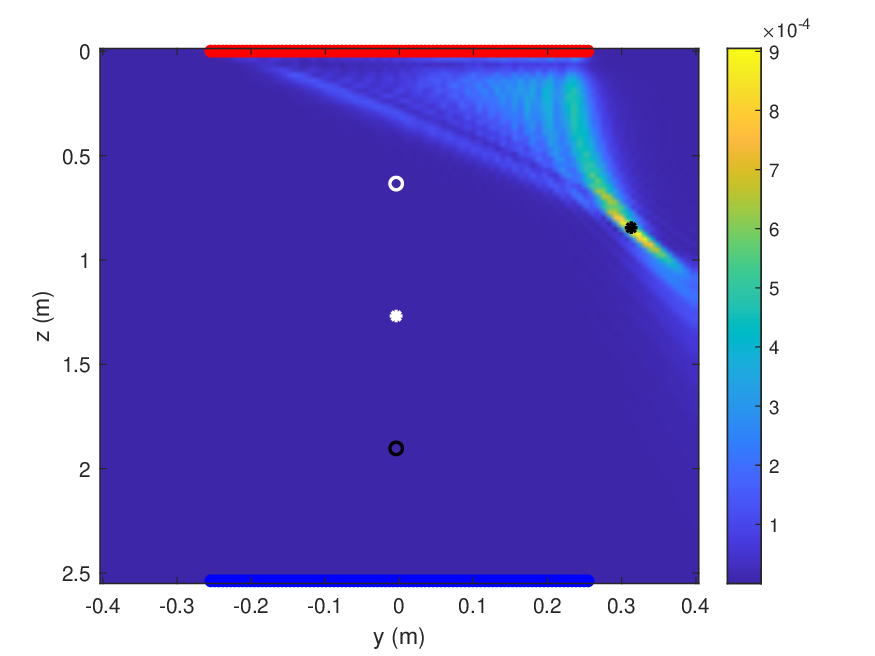}}
		\subfigure[Comparison of CRBs with $\Gamma = 0$ dB.]{ \label{fig:Target_CRB_comp}
			\includegraphics[width=1.7in]{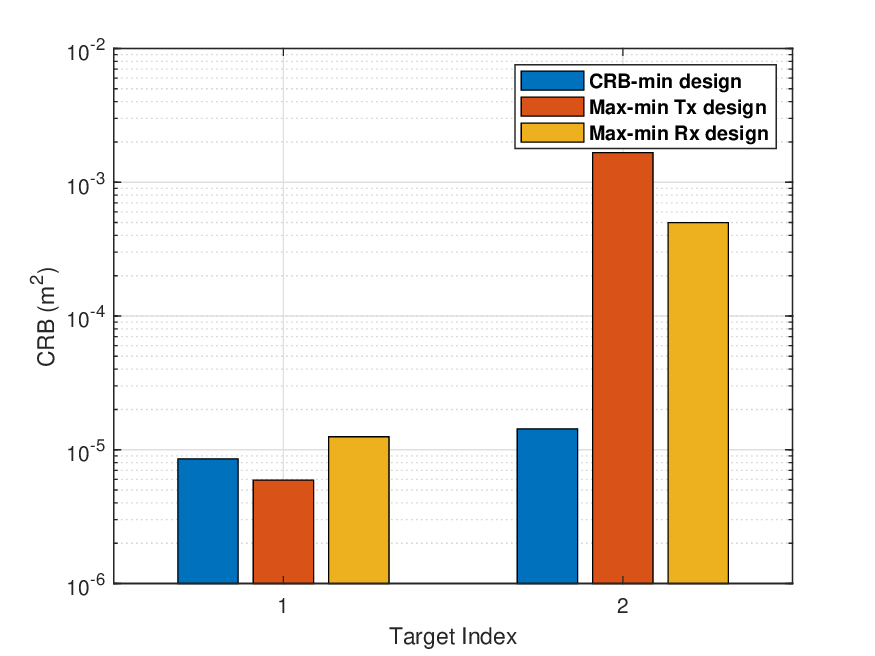}}
		\caption{(a)-(c) and (e)-(g). The comparison of the beampattern of CRB-min, Max-min Tx, and Max-min Rx designs with $\Gamma$ = 5 dB and 25 dB. The black circle and the black asterisk denote the first and the second target, respectively, while the white circle and the white asterisk denotes the first and the second CU, respectively. The red line represents the Tx array looking from the top while the blue line represents the Rx array. (d). Sum-CRB of two targets versus SINR requirement. (h). Minimum target illumination power or echo signal power versus SINR requirement. (l). The comparison of CRBs for both targets.}
		\label{fig:isac_multi_comp_beam}
\end{figure*}
In this subsection, we proceed to examine the general case with multiple targets and CUs. We first reveal the communication versus sensing tradeoffs and the beam focusing capabilities enabled by ELAA, and then validate the proposed design from the perspective of complexity reduction.


The setup is the same as the previous subsection except that we set $n_x = 16, n_y = 96$, and $D_z = 2.5371$ m.  
There exist two targets, the first one lying at $(0,0,3D_z/4)$, and the other at $(0,D_z/8,D_z/3)$ with their complex reflection coefficients being $b_1 = b_2 = 1$. There also exist two CUs, with the first one lying at $(0,0,D_z/4)$ and the second one at $(0,0,D_z/2)$. Only LoS paths between Tx and CUs are considered throughout this subsection. Furthermore, we set $\Gamma_u = \Gamma, \forall u \in \mathcal{U}$. Under the case with multiple targets, the design via (P2) is different from that via (P3) and thus we further use Max-min Tx to denote the design via (P2) and Max-min Rx to denote that via (P3).

Fig. \ref{fig:2Dbeam_CRB_isac_5db} - Fig. \ref{fig:2Dbeam_Rx_max_min_isac_5db}  and Fig. \ref{fig:2Dbeam_CRB_isac_25db} - \ref{fig:2Dbeam_Rx_max_min_isac_25db} show the beampattern of CRB-min, Max-min Tx, and Max-min Rx designs under the considered case with $\Gamma = 5$ dB and 25 dB, respectively.
Specifically, when the SINR requirement of CUs is low (e.g., $\Gamma = 5$ dB), the energy is mainly allocated towards both targets in black,
with one on the right and the other at the bottom. 
In particular, it is observed that a larger portion of energy is allocated towards the first target in black circle with Max-min Tx design in Fig. \ref{fig:2Dbeam_Tx_max_min_isac_5db} than that with Max-min Rx design in Fig. \ref{fig:2Dbeam_Rx_max_min_isac_5db}, showing the difference between Max-min Tx and Max-min Rx. By contrast, the CRB-min design allocates power in a more balanced way, as shown in Fig. \ref{fig:2Dbeam_CRB_isac_5db}. 
When the SINR requirement of CUs is high (e.g., $\Gamma = 25$ dB), it is observed that although some portions of energy are still allocated towards the locations of two targets to maintain the sensing performance, more energy is focused towards two CUs lying at the $z$-axis to meet the more stringent communication requirement in all the three designs shown in Fig. \ref{fig:2Dbeam_CRB_isac_25db}, Fig. \ref{fig:2Dbeam_Tx_max_min_isac_25db}, and Fig. \ref{fig:2Dbeam_Rx_max_min_isac_25db}. Furthermore, Fig. \ref{fig:2Dbeam_CRB_W1} - Fig. \ref{fig:2Dbeam_CRB_Rd} show the decomposed individual beampatterns of the beampattern in Fig. \ref{fig:2Dbeam_CRB_isac_25db}. It is observed that the beampattern by $\bm{W}_1$ is focused towards user 1 while eliminating the interference towards user 2 while the beampattern by $\bm{W}_2$ is focused towards user 2 while eliminating the interference towards user 1, which shows the superior beam focusing capability of ELAA in the near-field region. While the sensing performance of the first target is ensured by $\bm{W_1}$ and $\bm{W}_2$, the sensing performance of the second target is guaranteed through dedicated radar signal with $\bm{R}_d$, as shown in Fig. \ref{fig:2Dbeam_CRB_Rd}.

Besides, Fig. \ref{fig:isac_CRB_gamma} shows the tradeoff between multi-user communication and multi-target sensing in terms of sum-CRB, while Fig. \ref{fig:isac_mu_gamma} shows that in terms of minimum target illumination power or minimum target echo signal power, i.e., $\mu$ versus $\Gamma$. One observes that when $\Gamma$ increases, $\mu$ for Max-min Tx design and Max-min Rx design gradually decreases, which is intuitive. However, back to Fig. \ref{fig:isac_CRB_gamma}, the corresponding sum-CRB does not change in the same way as $\Gamma$ gradually increases, as their design objectives are different. These results again allude to the fact that optimizing the 3D positioning CRB is no longer equivalent to maximizing signal power or sensing SNR. Furthermore, it is observed from Fig. \ref{fig:isac_CRB_gamma} that when $\Gamma$ is low, the achieved sum-CRB by CRB-min design is same as that achieved by its counterpart without considering SINR constraints since the SINR constraints are not active. When $\Gamma$ increases, the sum-CRB degrades due to more stringent SINR requirements. Finally, Fig. \ref{fig:Target_CRB_comp} shows the comparison of CRBs for both targets achieved in all the three designs when the SINR constraints are not active. It is observed that CRB-min design significantly reduces the CRBs for both targets compared with the other two designs.

\begin{figure}[t]
		\centering
		\setlength{\abovecaptionskip}{+4mm}
		\setlength{\belowcaptionskip}{+1mm}
		\subfigure[Sum-CRB versus $n_y$.]{ \label{fig:CRB_n_y}
			\includegraphics[width=1.6375in]{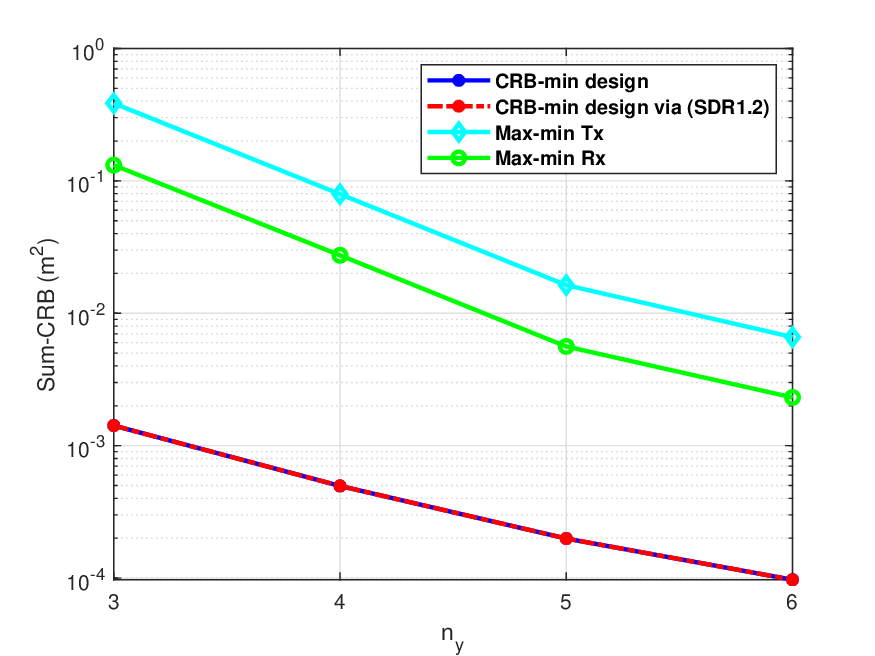}}
		\subfigure[Complexity versus $n_y$.]{ \label{fig:Complexity_n_y}
			\includegraphics[width=1.6375in]{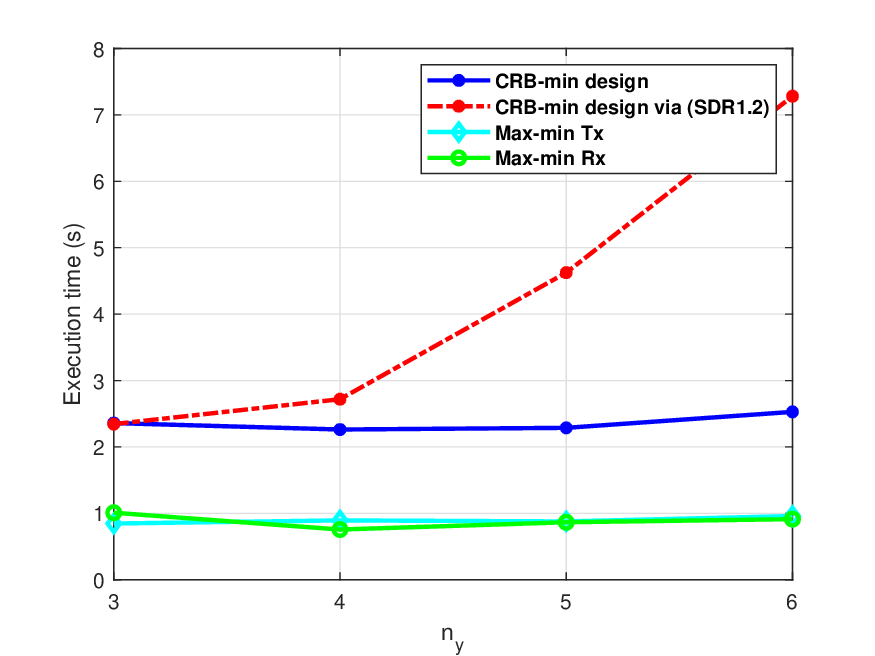}}
		\caption{(a). Sum-CRB of two targets versus $n_y$; and (b). The computational complexity in terms of execution time versus $n_y$.}
		\label{fig:num_N_complexity}
\end{figure}


Next, we show the importance of Proposition \ref{Prop:Rx_span} in significantly reducing the computational complexity in solving (P1.1), particularly under the case with ELAA. We run the simulation based on Intel(R) Core(TM) i7-10700F CPU @2.90GHz.
Specifically, we examine the computational time of different approaches, including the original CRB-min design based on directly solving (SDR1.2) and the equivalent CRB-min design utilizing the low-rank solution structure revealed in Propositio \ref{Prop:Rx_span}. We set $f_c = 1.5$ GHz, $\Gamma = 25$ dB, $n_x = 3$ with $n_y$ ranging from 3 to 6, i.e., 9 - 18 antennas at both Tx UPA and Rx UPA with half-wavelength spacing, and
$D_z = d_\text{nf}/20 = 1.205$ m. There exist two targets, i.e., $K = 2$, with one lying at $(0, 0, 3D_z/4)$ and the other at $(0,D_z/4,D_z/3)$, and their complex reflection coefficients equal to $b_1 = b_2 = 1$. There exists one CU, i.e., $U=1$, with its position at $(0, 0, D_z/4)$. The other setups are the same as those in the previous subsection. Notice that here we consider a relatively small number of antennas to facilitate comparison. This is due to the fact that when the number of total antennas at Tx or Rx is larger than 20, CRB-min design via directly solving (SDR1.2) is too slow due to the high computational complexity.

The results are shown in Fig. \ref{fig:num_N_complexity}. First, we show the sum-CRB versus $n_y$ in Fig. \ref{fig:CRB_n_y}, with fixed power budget $P_T$ and fixed SINR constraint for the single CU. 
One could observe that the sum-CRB obtained via directly solving (SDR1.2) and that via solving (SDR1.3) are the same, showing that the low-complexity approach based on the low-rank solution structure in Proposition \ref{Prop:Rx_span} yields exactly the same result as the original approach. However, as seen in Fig. \ref{fig:Complexity_n_y}, as the number of antennas increases, the execution time for CRB-min design based on solving (SDR1.2) directly becomes increasingly longer, while the execution time for the other three design approaches is much shorter.

\section{Conclusion}

This paper studied a near-field ISAC system equipped with ELAA, in which the BS deployed with enormous number of antennas transmits wireless signals to communicate with multiple downlink CUs while simultaneously using the echo signals to localize multiple point targets in the 3D space. To balance the sensing and communication performance tradeoff, we designed the transmit covariance matrix to optimize the localization performance while ensuring the SINR constraints at CUs. Specifically, we formulated three design problems by considering different 3D localization performance metrics, including minimizing the sum-CRB for estimating 3D locations, maximizing the minimum target illumination power, and maximizing the minimum target echo signal power. Their global optimal solutions were then obtained via SDR, together with the proof of their solution tightness. More importantly, it was rigorously shown that their optimal solutions have low-rank structures depending on sensing and communication channel matrices, which can be used to greatly reduce the complexity of the SDR-based solutions.
In the special case with a single collocated target/CU, we identified some special configurations where the optimal solutions to the proposed three designs and the SINR-maximization design become identical and admit a simple closed form.
Numerical results were presented, which are in accordance with the theoretical analysis, showing the benefits of the proposed three designs in optimizing both sensing and communication performance. It was also shown that in the collocated target/CU case, the optimal strategies for CRB-minimization and SINR-maximization are in general different and when the target/CU moves away from the transmitter/receiver, the CRB first decreases and then increases, which differ from the results in the conventional far-field setting. Besides, all the proposed designs with ELAAs demonstrated the capability of beam focusing and distance-domain spatial multiplexing, which can greatly enhance the performance of both multi-user communication and multi-target 3D localization. It was also revealed that by exploiting larger number of antennas and the scattered NLoS path from the target to the CU, better localization performance could be achieved with negligible loss in communication performance.


\appendix

\subsection{Proof of Proposition \ref{Prop:Rx_span}}\label{app:low_rank_structure}

As $\bm{I} = \br{P}_{\br{U}_{sc}}^{\perp} + \br{P}_{\br{U}_{sc}}$, where $\br{P}_{\br{U}_{sc}} = \br{U}_{sc} (\br{U}_{sc}^H \br{U}_{sc})^{-1} \br{U}_{sc}^H$ and $\br{P}_{\br{U}_{sc}}^{\perp}$ denote the projection operator onto the subspace and the orthogonal subspace of $\br{U}_{sc}$, respectively, we have
\begin{align}\label{eq:decomp_Rx_sc}
\bm{R}_X^* & = \sum_{u=1}^{U} \bm{W}_{u}^* + \bm{R}_d^* \\
\nonumber
& = \br{P}_{\br{U}_{sc}} (\sum_{u=1}^{U} \bm{W}_{u}^* + \bm{R}_d^*) \br{P}_{\br{U}_{sc}} + \sum_{u=1}^{U} \tilde{\bm{W}}_u^* + \tilde{\bm{R}}_d^*,
\end{align} 
where for any $u \in \mathcal{U}$,
\begin{align}
\nonumber
\tilde{\bm{W}}_u^* & =  \br{P}_{\br{U}_{sc}}^{\perp} \bm{W}_{u}^* \br{P}_{\br{U}_{sc}}^{\perp} + \br{P}_{\br{U}_{sc}} \bm{W}_{u}^* \br{P}_{\br{U}_{sc}}^{\perp} + \br{P}_{\br{U}_{sc}}^{\perp} \bm{W}_{u}^* \br{P}_{\br{U}_{sc}}, \\
\nonumber
\tilde{\bm{R}}_d^* & = \br{P}_{\br{U}_{sc}}^{\perp} \bm{R}_d^* \br{P}_{\br{U}_{sc}}^{\perp} + \br{P}_{\br{U}_{sc}} \bm{R}_d^* \br{P}_{\br{U}_{sc}}^{\perp} + \br{P}_{\br{U}_{sc}}^{\perp} \bm{R}_d^* \br{P}_{\br{U}_{sc}}.
\end{align}
Let $\tilde{\bm{R}}_X^* \triangleq \sum_{u=1}^{U} \tilde{\bm{W}}_u^* + \tilde{\bm{R}}_d^*$. First, for the constraint in (\ref{eq:CRB_ISAC_P2r_Fish}), one can readily verify that 
\begin{align}\label{eq:FIM_null_sc}
\nonumber
& \bm{V}^H \tilde{\bm{R}}_X^* \bm{V} =  \bm{V}^H \tilde{\bm{R}}_X^* \dot{\bm{V}}_{\br{x}} = \dot{\bm{V}}_{\br{x}}^H \tilde{\bm{R}}_X^* \bm{V} = \dot{\bm{V}}_{\br{x}}^H \tilde{\bm{R}}_X^* \dot{\bm{V}}_{\br{x}} = \bm{0}, \\
\nonumber
& \bm{V}^H \tilde{\bm{R}}_X^* \dot{\bm{V}}_{\br{y}} = \dot{\bm{V}}_{\br{y}}^H \tilde{\bm{R}}_X^* \bm{V} = \dot{\bm{V}}_{\br{y}}^H \tilde{\bm{R}}_X^* \dot{\bm{V}}_{\br{y}} = \bm{0}, \\
\nonumber
& \bm{V}^H \tilde{\bm{R}}_X^* \dot{\bm{V}}_{\br{z}} = \dot{\bm{V}}_{\br{z}}^H \tilde{\bm{R}}_X^* \bm{V} = \dot{\bm{V}}_{\br{z}}^H \tilde{\bm{R}}_X^* \dot{\bm{V}}_{\br{z}} = \bm{0}, \\
& \dot{\bm{V}}_{\br{x}}^H \tilde{\bm{R}}_X^* \dot{\bm{V}}_{\br{y}} = \dot{\bm{V}}_{\br{x}}^H \tilde{\bm{R}}_X^* \dot{\bm{V}}_{\br{z}} = \dot{\bm{V}}_{\br{y}}^H \tilde{\bm{R}}_X^* \dot{\bm{V}}_{\br{z}} = \bm{0}.
\end{align}
By checking the expression for each block in $\br{F}$ in Proposition \ref{Pro:F_deri}, one note that the FIM does not depend on $\tilde{\bm{R}}_X^*$. 
Second, for the SINR constraint in (\ref{eq:CRB_ISAC_P2r_SINR}), we have
\begin{align}
\label{eq:SINR_null_Wp}
&\text{tr}\left(\bm{h}_u \bm{h}_u^H \tilde{\bm{W}}_k\right) = \bm{h}_u^H \tilde{\bm{W}}_k \bm{h}_u = 0, \enspace \forall k,u \in \mathcal{U}, \\
\label{eq:SINR_null_Rd}
&\text{tr}\left(\bm{h}_u \bm{h}_u^H \tilde{\bm{R}}_d\right) = \bm{h}_u^H \tilde{\bm{R}}_d \bm{h}_u = 0, \enspace \forall u \in \mathcal{U}.
\end{align}
Third, for the sum-power constraint in (\ref{eq:CRB_ISAC_P2r_power}), we decompose $\bm{R}_X^* = \br{\delta} \br{\delta}^H$; accordingly, we have
\begin{align}
\nonumber
\operatorname{tr}(\tilde{\bm{R}}_X^*) 
& = \operatorname{tr}((\br{P}_{\br{U}_{sc}}^{\perp} + \br{P}_{\br{U}_{sc}}) \br{\delta} \br{\delta}^H \br{P}_{\br{U}_{sc}}^{\perp} + \br{P}_{\br{U}_{sc}}^{\perp} \br{\delta} \br{\delta}^H \br{P}_{\br{U}_{sc}})\\
\label{eq:trace_zero_cond}
& = \operatorname{tr}(\br{P}_{\br{U}_{sc}}^{\perp} \br{\delta} \br{\delta}^H \br{P}_{\br{U}_{sc}}^{\perp}) = \|\br{P}_{\br{U}_{sc}}^{\perp} \br{\delta}\|_F^2 \geq 0,
\end{align}
where the equality holds if and only if (iff) $\br{P}_{\br{U}_{sc}}^{\perp} \bm{\delta} = \br{0}$, implying $\tilde{\bm{R}}_X^* = \br{0}$. 

Now, suppose that for the optimal $\bm{R}_X^*$, its corresponding $\tilde{\bm{R}}_X^* \neq \br{0}$; thus, we have $\operatorname{tr}(\tilde{\bm{R}}_X^*) > 0$ according to the argument in (\ref{eq:trace_zero_cond}). If we set $\{\tilde{\bm{W}}_u^*\}$ and $\tilde{\bm{R}}_d^*$ all equal to zero matrices, this will not alter the FIM according to (\ref{eq:FIM_null_sc}). Furthermore, according to (\ref{eq:SINR_null_Wp}) and (\ref{eq:SINR_null_Rd}), the corresponding SINR value also remains unchanged. However, we now have remaining power budget.
If we scale up $\{\br{P}_{\br{U}_{sc}}\bm{W}_u^*\br{P}_{\br{U}_{sc}}\}$ and $\br{P}_{\br{U}_{sc}}\bm{R}_d^*\br{P}_{\br{U}_{sc}}$ proportionally in (\ref{eq:decomp_Rx_sc}) to consume the remaining power budget, this will generate a new set of solution, in which we have a scaling-up FIM and accordingly a smaller CRB. Besides, the SINR value will also increase and still meet the SINR requirements. As a result, we find a new set of feasible solutions with smaller CRB, leading to an immediate contradiction to the assumption that $\bm{R}_X^*$ is optimal. Hence, we have $\tilde{\bm{R}}_X^* = \bm{0}$ and the optimal $\bm{R}_X$ that satisfies $\bm{R}_X^* = \br{P}_{\br{U}_{sc}} \br{\delta} \br{\delta}^H \br{P}_{\br{U}_{sc}} = \br{U}_{sc} \br{\Sigma}_{sc} \br{U}_{sc}^H$ with $\br{\Sigma}_{sc} = (\br{U}_{sc}^H \br{U}_{sc})^{-1} \br{U}_{sc}^H \br{\delta} \br{\delta}^H \br{U}_{sc} (\br{U}_{sc}^H \br{U}_{sc})^{-1}$, which is a positive semi-definite matrix. 

We proceed to prove that $\tilde{\bm{W}}_1^* = ... = \tilde{\bm{W}}_U^* = \tilde{\bm{R}}_d^* = \bm{0}$. We first decompose $\bm{R}_d^*$ and $\{\bm{W}_u^*\}_{u=1}^U$ into
\begin{align}
\bm{W}_u^* = \br{\delta}_u \br{\delta}_u^H, \enspace u \in \mathcal{U}, \enspace \bm{R}_d^* = \br{\delta}_d \br{\delta}_d^H.
\end{align}
With $\tilde{\bm{R}}_X^* =  \sum_{u=1}^{U} \tilde{\bm{W}}_u^* + \tilde{\bm{R}}_d^* = \bm{0}$, we have 
\begin{align}\label{eq:trace_ud_zero}
& \operatorname{tr}(\tilde{\bm{R}}_X^*)  = \operatorname{tr}(\br{P}_{\br{U}_{sc}}^{\perp} (\sum_{u=1}^{U} \br{\delta}_u \br{\delta}_u^H + \br{\delta}_d \br{\delta}_d^H) \br{P}_{\br{U}_{sc}}^{\perp}) \\
\nonumber
& = \sum_{u=1}^U \|\br{P}_{\br{U}_{sc}}^{\perp} \br{\delta}_u\|_F^2 + \|\br{P}_{\br{U}_{sc}}^{\perp} \br{\delta}_d\|_F^2 = 0,
\end{align}
which implies that $\br{P}_{\br{U}_{sc}}^{\perp} \br{\delta}_u = \bm{0}, \enspace \forall u \in \mathcal{U},$ and $\br{P}_{\br{U}_{sc}}^{\perp} \br{\delta}_d = \bm{0}$, 
leading to $\tilde{\bm{W}}_1^* = ... = \tilde{\bm{W}}_U^* = \tilde{\bm{R}}_d^* = \bm{0}$. As such, we have
\begin{align}
\bm{W}_u^*  = \br{U}_{sc} \br{\Sigma}_u \br{U}_{sc}^H, \enspace u \in \mathcal{U}, \enspace
\bm{R}_d^*  = \br{U}_{sc} \br{\Sigma}_d \br{U}_{sc}^H,
\end{align}
with $\br{\Sigma}_{u} = (\br{U}_{sc}^H \br{U}_{sc})^{-1} \br{U}_{sc}^H \br{\delta}_u \br{\delta}_u^H \br{U}_{sc} (\br{U}_{sc}^H \br{U}_{sc})^{-1}, u \in \mathcal{U}$,
and $\br{\Sigma}_{d} = (\br{U}_{sc}^H \br{U}_{sc})^{-1} \br{U}_{sc}^H \br{\delta}_d \br{\delta}_d^H \br{U}_{sc} (\br{U}_{sc}^H \br{U}_{sc})^{-1}$. 

\subsection{Proof of Lemma \ref{lemma:w_s_range_RXs}}\label{proof:ws_in_R_Xs}

We prove the lemma via contradiction. Suppose that $\bm{w}^s$ does not lie in $\mathcal{R}(\br{U})$. Then $\bm{w}^s$ can be expressed as $\bm{w}^s = \br{U} \bm{x} + \br{U}_n \bm{y}, \bm{y} \neq \bm{0}$, where $\mathcal{R}(\br{U}_n) = \mathcal{R}^\perp(\br{U})$. Thus, we have
\begin{align}
\nonumber
\bm{R}_d^s & = \bm{R}_X^s - \bm{w}^s (\bm{w}^s)^H = \br{U} (\bm{\Sigma}_s - \bm{x}\bm{x}^H) \br{U}^H \\
&  - \br{U}_n \bm{y} \bm{x}^H \br{U}^H - \br{U} \bm{x} \bm{y}^H \br{U}_n^H  - \br{U}_n \bm{y} \bm{y}^H \br{U}_n^H.
\end{align} 
Since $\bm{y} \neq \bm{0}$, by defining $\bm{t} = \br{U}_n \bm{y} \neq \bm{0}$, we have $\bm{t}^H \bm{R}_d^s \bm{t} = - \|\bm{y}\|^4 < 0$. This violates the positive semi-definite constraint of $\bm{R}_d^s$. Thus, $\bm{w}^s$ lies in the range space of $\bm{R}_X^s$.

\subsection{Proof of Proposition \ref{prop:CRB_min_MRC}}\label{proof:single_special_merged_Pc_Ps}

With the single collocated target/CU, we define
\begin{align}\label{eq:VX_partial_single}
\dot{\bm{A}_{\bm{\mathrm{u}}}} & = \frac{\partial \bm{a}(\bm{\mathrm{l}}_1)}{\partial \mathrm{u}_1} \triangleq \dot{\bm{a}}_\mr{u}, \enspace \mr{u} \in \{\mr{x},\mr{y},\mr{z}\}, \enspace \bm{A} \triangleq \bm{a},  \\
\dot{\bm{V}_{\bm{\mathrm{u}}}} & = \frac{\partial \bm{v}(\bm{\mathrm{l}}_1)}{\partial \mathrm{u}_1} \triangleq \dot{\bm{v}}_\mr{u}, \enspace \mr{u} \in \{\mr{x},\mr{y},\mr{z}\}, \enspace \bm{V} \triangleq \bm{v}.
\end{align}
Accordingly, the following equations hold due to the symmetric configuration in Fig. \ref{fig:CRB_d0_single_setup_isac}:
\begin{align}
\nonumber
& \dot{\bm{a}}_\mathrm{x}^H \bm{a} = \bm{a}^H \dot{\bm{a}}_\mathrm{x} = \dot{\bm{v}}_\mathrm{x}^H \bm{v} = \bm{v}^H \dot{\bm{v}}_\mathrm{x} =0, \\
\label{eq:dxyz_ortho_isac}
& \dot{\bm{a}}_\mathrm{y}^H \bm{a} = \bm{a}^H \dot{\bm{a}}_\mathrm{y} = \dot{\bm{v}}_\mathrm{y}^H \bm{v} = \bm{v}^H \dot{\bm{v}}_\mathrm{y} =0, \\
\nonumber
& \dot{\bm{a}}_\mathrm{x}^H \dot{\bm{a}}_\mathrm{y} = \dot{\bm{a}}_\mathrm{x}^H \dot{\bm{a}}_\mathrm{z} = \dot{\bm{a}}_\mathrm{y}^H \dot{\bm{a}}_\mathrm{z} = 0, \\
\label{eq:vxyz_ortho_isac}
&\dot{\bm{v}}_\mathrm{x}^H \dot{\bm{v}}_\mathrm{y} = \dot{\bm{v}}_\mathrm{x}^H \dot{\bm{v}}_\mathrm{z} = \dot{\bm{v}}_\mathrm{y}^H \dot{\bm{v}}_\mathrm{z} = 0, \enspace \|\bm{a}\|^2   =  \|\bm{v}\|^2,\\
&  \|\dot{\bm{a}}_\mathrm{x}\|^2 = \|\dot{\bm{v}}_\mathrm{x}\|^2 = \|\dot{\bm{a}}_\mathrm{y}\|^2 = \|\dot{\bm{v}}_\mathrm{y}\|^2, \enspace \|\dot{\bm{a}}_\mathrm{z}\|^2 = \|\dot{\bm{v}}_\mathrm{z}\|^2.
\label{eq:a_norm_pds_partial_isac}
\end{align}
For the bistatic sensing setup in Fig. \ref{fig:CRB_d0_single_setup_isac}(a), we have
\begin{align}
& \dot{\bm{a}}_\mathrm{z}^H \bm{a}   = -\dot{\bm{v}}_\mathrm{z}^H \bm{v} =  \overline{\bm{a}^H \dot{\bm{a}}_\mathrm{z}} = -\overline{\bm{v}^H \dot{\bm{v}}_\mathrm{z}},
\end{align}
while for the monostatic sensing setup in Fig. \ref{fig:CRB_d0_single_setup_isac}(b), we have
\begin{align}
& \dot{\bm{a}}_\mathrm{z}^H \bm{a}   = \dot{\bm{v}}_\mathrm{z}^H \bm{v} =  \overline{\bm{a}^H \dot{\bm{a}}_\mathrm{z}} = \overline{\bm{v}^H \dot{\bm{v}}_\mathrm{z}}.
\end{align}
Since $\bm{R}_X^s = \br{U}_{r}^* \br{\Sigma}_{r}^* \br{U}_{r}^T$, to facilitate analysis, we perform the QR decomposition of $\br{U}_{r}$ as $\br{U}_{r} = \br{Q}_{r}\br{R}_{r}$. As a result, $(\bm{R}_X^s)^* =  \br{Q}_{r} \br{R}_{r} \br{\Sigma}_{r} \br{R}_{r}^H \br{Q}_{r}^H$, and we have
\begin{align}
\nonumber
(\bm{R}_X^s)^* = \br{Q}_{r} \underbrace{\left[
	\arraycolsep=1.2pt\def\arraystretch{1.0}
	\begin{array}{cccc}
		x_1 & x_{12} & x_{13} & x_{14} \\[1pt]
		x_{12}^* & x_2 & x_{23} & x_{24}  \\[1pt]
		x_{13}^* & x_{23}^* & x_3 & x_{34}  \\[1pt]
		x_{14}^* & x_{24}^* & x_{34}^* & x_4 
	\end{array}\right]}_{\tilde{\bm{\Sigma}}_{\mr{r}} \triangleq \br{R}_{r} \br{\Sigma}_{r} \br{R}_{r}^H} \br{Q}_{r}^H, \enspace x_i \geq 0, i=1,2,3,4,
	\end{align}
	where $\br{Q}_{r} = \left[\frac{\bm{v}}{\|\bm{v}\|}, \frac{\dot{\bm{v}}_\mr{x}}{\|\dot{\bm{v}}_\mr{x}\|},\frac{\dot{\bm{v}}_\mr{y}}{\|\dot{\bm{v}}_\mr{y}\|},\frac{\bm{\omega}}{\|\bm{\omega}\|}\right]$, and $\br{Q}_{r}^H \br{Q}_{r} = \bm{I}_4$
	due to the orthogonality conditions in (\ref{eq:dxyz_ortho_isac}) and (\ref{eq:vxyz_ortho_isac}).
	Besides, we have
	\begin{align}\label{eq:relation_w_vz}
\bm{\omega} = \frac{\dot{\bm{v}}_\mr{z}}{\|\dot{\bm{v}}_\mr{z}\|} - \frac{\bm{v}^H \dot{\bm{v}}_\mr{z}}{\|\bm{v}\|^2 \|\dot{\bm{v}}_\mr{z}\|} \bm{v},
\end{align} 
which is unique, to within scaling by any number of unit modulus.	Based on (\ref{eq:relation_w_vz}), one can show that
\begin{align}\label{eq:Relation_w_vz}
\|\dot{\bm{v}}_\mr{z}\|^2 \|\bm{v}\|^2 - |\dot{\bm{v}}_\mr{z}^H \bm{v}|^2 = \|\bm{v}\|^2 \frac{|\dot{\bm{v}}_\mr{z}^H \bm{\omega}|^2}{\|\bm{\omega}\|^2}.
\end{align}
In our considered case, $\br{F}$ in Proposition \ref{Pro:F_deri} is recast as
\begin{align}\label{eq:Schur_DF}
\bm{\mathrm{F}} =\left[\begin{array}{cc}
	\bm{G} & \bm{H} \\
	\bm{H}^T & \bm{R}
\end{array}\right],
\end{align}
where $\bm{G} = \bm{\mathrm{F}}([1:3],[1:3]), \bm{R} = \bm{\mathrm{F}}([4:5],[4:5])$, and $\bm{H} = \bm{\mathrm{F}}([1:3],[4:5])$. The Schur complement of block $\bm{R}$ of $\bm{\mathrm{F}}$ (the equivalent FIM) is expressed as
\begin{align}
\br{D} = \bm{G} - \bm{H}\bm{R}^{-1}\bm{H}^T \in \mathbb{R}^{3 \times 3},
\end{align}
with $\br{D}^{-1} = \left[\bm{\mathrm{F}}^{-1}\right]_{3 \times 3}$ according to the Matrix Inversion Lemma\cite{horn2012matrix}. As $\br{F} \succ \bm{0}$, $\br{D} \succ \bm{0}$, $\tilde{\br{D}} \triangleq \xi \br{D} \succ \bm{0}$, where $\xi \triangleq \frac{\sigma^2}{2 |b|^2 L}$, the CRB of each coordinate is then expressed as
\begin{align}
\label{eq:CRB_x}
\text{CRB}_\mathrm{x} & = \left[\bm{\mathrm{F}}^{-1}\right]_{1,1}=\xi(\tilde{d}_{22}\tilde{d}_{33}-\tilde{d}_{23}^2)/|\tilde{\br{D}}|>0,\\
\text{CRB}_\mathrm{y} & = \left[\bm{\mathrm{F}}^{-1}\right]_{2,2}=\xi(\tilde{d}_{11}\tilde{d}_{33}-\tilde{d}_{13}^2)/|\tilde{\br{D}}|>0,\\
\text{CRB}_\mathrm{z} & = \left[\bm{\mathrm{F}}^{-1}\right]_{3,3}=\xi(\tilde{d}_{11}\tilde{d}_{22}-\tilde{d}_{12}^2)/|\tilde{\br{D}}|>0,
\end{align}	
where $\tilde{d}_{ij}$ denotes the $(i,j)$th entry of $\tilde{\br{D}}$.		
Substituting (\ref{eq:dxyz_ortho_isac}) - (\ref{eq:Relation_w_vz}) into the definition of the FIM $\br{F} \in \mathbb{R}^{5 \times 5}$  in
Proposition \ref{Pro:F_deri} and the definition of the sum-CRB in (\ref{eq:CRB_position_isac}) and (\ref{eq:Sum_CRB}), through some mathematical manipulation, it follows that
\begin{align}
\tilde{d}_{11} & = \|\bm{v}\|^2 \|\dot{\bm{v}}_\mr{x}\|^2 (x_1+x_2-\frac{|x_{12}|^2}{x_1}), \\
\tilde{d}_{22} & = \|\bm{v}\|^2 \|\dot{\bm{v}}_\mr{y}\|^2 (x_1+x_3-\frac{|x_{13}|^2}{x_1}), \\
\tilde{d}_{33} & = (\|\bm{v}\|^2 \|\dot{\bm{v}}_\mr{z}\|^2 - |\dot{\bm{v}}_\mr{z}^H \bm{v}|^2) (x_1+x_4-\frac{|x_{14}|^2}{x_1}),\\
\tilde{d}_{12} & = \|\bm{v}\|^2 \|\dot{\bm{v}}_\mr{x}\|^2 \mathfrak{R} \{x_{23} - \frac{x_{12}^*}{x_1} x_{13} \},\\
\tilde{d}_{13} & =   \|\bm{v}\|^2 \|\dot{\bm{v}}_\mr{x}\|/\|\bm{\omega}\| \mathfrak{R} \{x_{24} \bm{\omega}^H \dot{\bm{v}}_\mr{z} - \frac{x_{12}^*}{x_1} x_{14} \dot{\bm{v}}_\mr{z}^H \bm{\omega}\},
\end{align}
\begin{align}
\tilde{d}_{23} & =   \|\bm{v}\|^2 \|\dot{\bm{v}}_\mr{y}\|/\|\bm{\omega}\| \mathfrak{R} \{x_{34} \bm{\omega}^H \dot{\bm{v}}_\mr{z} - \frac{x_{13}^*}{x_1} x_{14} \dot{\bm{v}}_\mr{z}^H \bm{\omega}\}, 
\end{align}
where the above simplified expressions hold for both the monostatic and bistatic sensing setups in Fig. \ref{fig:CRB_d0_single_setup_isac}.
Based on (\ref{eq:CRB_x}), the CRB for $\mr{x}$-coordinate is alternatively expressed as
\begin{align}
\text{CRB}_\mathrm{x} =\xi \frac{1}{\tilde{d}_{11}+\frac{2 \tilde{d}_{12} \tilde{d}_{13} \tilde{d}_{23} - \tilde{d}_{22} \tilde{d}_{13}^2 - \tilde{d}_{33} \tilde{d}_{12}^2}{\tilde{d}_{22}\tilde{d}_{33}-\tilde{d}_{23}^2}}.
\end{align}
According to Fischer's inequality \cite{horn2012matrix}, we have
\begin{align}
|\tilde{\br{D}}|  \leq \tilde{d}_{11} & (\tilde{d}_{22}  \tilde{d}_{33} - \tilde{d}_{23}^2) \Rightarrow  \\
\nonumber
& 2 \tilde{d}_{12} \tilde{d}_{13} \tilde{d}_{23} - \tilde{d}_{22} \tilde{d}_{13}^2 - \tilde{d}_{33} \tilde{d}_{12}^2 \leq 0,
\end{align}
where inequality holds iff $\tilde{d}_{12} = \tilde{d}_{13} = \tilde{d}_{23} = 0$, i.e., $\tilde{\br{D}}$ is diagonal.
It is easy to see that for any feasible solution $\tilde{\br{\Sigma}}_{\mr{r}}$, setting all of its non-diagonal elements equal to zero will increase the first term $\tilde{d}_{11}$ in the denominator of $\text{CRB}_{\mr{x}}$ as well as the second term to zero, thus resulting in lower $\text{CRB}_{\mr{x}}$. Similar results also hold for $\text{CRB}_{\mr{y}}$ and $\text{CRB}_{\mr{z}}$. In other words, for any optimal solution to (P1.s) under the considered collocated target/CU case without SINR constraints, the structure of $\tilde{\bm{\Sigma}}_{\mr{r}}$ must be diagonal and is expressed as $\tilde{\bm{\Sigma}}_{\mr{r}} = \operatorname{diag}(x_1,x_2,x_3,x_4)$. Thus, the CRB is simplified as
\begin{align}
	\nonumber
	& \text{CRB}(x_1,...,x_4)  = \underbrace{\frac{\xi}{(\|\dot{\bm{v}}_\mr{z}\|^2 \|\bm{v}\|^2 - |\dot{\bm{v}}_\mr{z}^H \bm{v}|^2) (x_1 + x_4) }}_{\text{CRB}_\mr{z}}   \\
	& + \underbrace{\frac{\xi}{\|\dot{\bm{v}}_\mr{y}\|^2 \|\bm{v}\|^2 (x_1+x_3)}}_{\text{CRB}_\mr{y}} + \underbrace{\frac{\xi}{\|\dot{\bm{v}}_\mr{x}\|^2 \|\bm{v}\|^2 (x_1+x_2) }}_{\text{CRB}_\mr{x}}.
\end{align}
Thus, (P1.s) without SINR constraint is simplified as 
\begin{align}\label{eq:CRB_sensing_Rx_diag_sim}
	\min_{x_1,...,x_4} \text{ CRB}(x_1,...,x_4)  \quad  \text{s.t. } \sum_{i=1}^{4} x_i \leq P_T.
\end{align}
We can see that concentrating power on $x_1$ will minimize $\text{CRB}_\mr{x}$, $\text{CRB}_\mr{y}$, and $\text{CRB}_\mr{z}$ altogether, leading to $x_1 = P_T, x_2=x_3=x_4=0$ or $\bm{R}_X^s = \frac{P_T}{\|\bm{v}\|^2} \bm{v}^* \bm{v}^T$ equivalently. The proof of Proposition \ref{prop:CRB_min_MRC} is thus complete.

\bibliographystyle{IEEEtran}

\bibliography{refsv3}

\end{document}